# Heterogeneity among migrants, education-occupation mis-match and returns to education: Evidence from India

Shweta Bahl

Indian Institute of Management Rohtak

Ajay Sharma

Indian Institute of Management Indore

## Abstract

*Using nationally representative data for India, this paper examines the incidence of education-occupation mismatch (EOM) and returns to education and EOM for internal migrants while considering the heterogeneity among them. In particular, this study considers heterogeneity arising because of the reason to migrate, demographic characteristics, spatial factors, migration experience, and type of migration. The analysis reveals that there exists variation in the incidence and returns to EOM depending on the reason to migrate, demographic characteristics, and spatial factors. The study highlights the need of focusing on EOM to increase the productivity benefits of migration. It also provides the framework for minimizing migrants' likelihood of being mismatched while maximizing their returns to education.*

**Acknowledgement:** We are thankful to the seminar participants at UNU-WIDER Conference 2019 and ICEF 2020 for their helpful comments.

***Correspondence Address***: Shweta Bahl, Indian Institute of Management Rohtak, Rohtak, Haryana (India)- 124010; e-mail: shweta.bahl28@gmail.com; f15shwetag@iimidr.ac.in.

**Ajay Sharma**, Indian Institute of Management Indore, Rau-Pithampur Road, Indore, (M.P.) India-453556; e-mail: ajays@iimidr.ac.in ; ajaysharma87@gmail.com. Phone: +91-731-2439622.

# Heterogeneity among migrants, education-occupation mismatch and returns to education: Evidence from India

## 1. Introduction

Migrants comprise of those who experience either a permanent or semi-permanent change of residence (Lee, 1966). There are several models that predict and explain migration as a result of investment in human capital. By now, economists and demographers have a clear-cut answer to why people migrate. Migration prompts numerous benefits. Migration reduces poverty, increases income, enhances upward mobility, escalates savings and assets formation, fosters investment in human capital, reduces vulnerability, improves food security, stimulates land markets at origin, increases local wages and demand for local goods and services, improves the economy, and tightens the rural labour markets (Bird and Deshingkar, 2013). Due to the costly nature of international migration, internal migration has a considerably large role in the process of upward mobility, development and economic growth. Internal migrants are defined as those who relocate within their home country, but outside their usual place of residence for a considerable duration.[1]

Internal migration is an important mechanism by which labor resources get redistributed nationally on account of changing demographic and economic forces (Greenwood, 1997). Various researchers have studied the patterns of internal migration and found that individuals choose to migrate toward regions that pay higher income (Borjas et al., 1992) and have low unemployment rate (Herzog et al., 1993). Therefore, individuals by being spatially flexible attempt to maximize their earnings and lower their chances of being unemployed. However, the extent to which migrants are able to efficiently match their education with their occupation and the impact of mismatch on their income is still a debatable question. This issue is interesting to examine because in a given occupation, productivity (Tsang and Levin, 1985) and job satisfaction (McGuinness and Sloane, 2011) is higher in a case where education required by an occupation matches the education attained by a worker as compared to a situation of mismatch. Further, migration has been considered as an important channel for achieving the highest level of productivity (Maxwell, 1988). Therefore, the exploration of the interlinkages between migration and education-occupation mismatch (EOM)[2] addresses the broader question of workers' ability to better utilize their human capital endowments by being spatially flexible. It also highlights the present state of the use of human capital endowments of internal migrants with respect to recipient regions. Moreover, since EOM has considerable consequences on income (Duncan and Hoffman, 1981; Verdugo and Verdugo, 1989), it can serve as an imperative indicator of migrants' success or failure at the destination. This paper, in general, attempts to contribute to this literature by analyzing the extent of and returns to EOM for migrants.

---

[1] In this study, an internal migrant is defined based on their current place of residence (a village, town or city) that is different from their last usual place of residence at the time of the survey, and the duration of such mobility is at least six months or more (which is the cutoff for domicile status identification) within the administrative boundaries of a country (National Sample Survey Office, 2010). This study focuses on India, and hence, the definition provided comes from the Census of India and National Sample Survey Office (NSSO). The definition is also used by the Government of India for administrative and public policy purposes. Other countries also use similar definitions with slight variation in the duration considered.

The last decade has witnessed growing contributions from researchers toward analyzing the relationship between migration and EOM. The current research on this issue has touched on two themes. The first segment explores the impact of migration on the likelihood of mismatch between education and occupation (henceforth, mismatched). This theme has further evolved in two strands. The first strand deals with international migration. Researchers have found that international migration leads to higher likelihood of being mismatched (Aleksynska and Tritah, 2013; Dahlstedt, 2011; Nielsen, 2011; Wald and Fang, 2008) and bases this finding on the imperfect transferability of human capital (Nieto et al., 2015). The second strand considers internal migration. Researchers have hypothesized that internal migration should lead to lower likelihood of being mismatched. This is because spatial flexibility leads to better access to jobs and consequently widens the adequate job opportunities available to a worker, which is grounded on Frank's (1978) theory of differential overqualification. The central aspect of this theory is that the incidence of EOM would be higher for workers with relative spatial inflexibility (Büchel and van Ham, 2003). Besides, the problem of human capital transferability is of limited concern for internal migration due to similar culture, language, infrastructure, and so on. Therefore, EOM may arise due to spatially constrained job-search and could be reduced by being spatially flexible within a home country.[3] While many studies have supported this argument by observing a higher likelihood of mismatch among natives than migrants, others have argued that there exists no significant relationship between migration and probability of being mismatched (Hensen et al., 2009; Iammarino and Marinelli, 2015; Jauhiainen, 2011).

---

[2] EOM refers to the incongruity between the attained education of a worker and the required education by her occupation (Duncan and Hoffman, 1981).

[3] Although this phenomenon is also present in the case of international migration, it is suppressed by the negative impact of imperfect portability of human capital.

The second segment analyzes the impact of migration on the returns to EOM. In the case of international migration, the literature has observed that migrants lose much more from not being correctly matched than natives do (Joona et al., 2014; Neilsen, 2011). While the literature on the first segment is quite rich, the literature on the second segment is scant in case of international migrants and almost non-existent for internal migrants. Further, apart from the international and internal migration, the existing literature has also examined the differential impact of the likelihood of being mismatched by source region (Joona et al., 2014), destination region (Iammarino and Marinelli, 2015), source-destination pairs (Aleksynska and Tritah, 2013), and education level (Croce and Ghignoni, 2015) of the migrants. Therefore, barring a few studies, others have considered migrants as a homogeneous group which could be misleading. This is because migrants can have different socio-economic, demographic, and other personal characteristics, which in turn, can have a differential impact on their labor market outcomes.

In order to address this concern, this study expands the theoretical model developed by Simpson (1992) and adapted by Büchel and van Ham (2003). The model relates a person's match status with the range of job opportunities. A person looking for a job in a particular local labor market has three options in the scenario of not finding an adequate job; first, not to accept any job (unemployment); second, settle for a mismatched job; third, migrate or commute to another labor market for better prospects. Based on the model, Büchel and van Ham (2003) observed that individuals who can widen the size of their labor market have a better chance of getting an adequately matched job. This paper argues that a series of decisions do not end here. Once an individual decides to migrate, there are other decisions that an individual has to take regarding location, type of migration, and so on. These decisions again depend on various factors such as availability of job opportunities at home location, assimilation, infrastructure facilities, and so on. Since the motivation to migrate varies for different choices, it raises an interesting question regarding the differences in the labor market outcomes, in general, and EOM, in particular, with respect to different kinds of migration.

This paper aims to answer this question, and therefore, estimates the differential returns to EOM while distinguishing the migrants by (i) reason to migrate, particularly, in search for a job or to take up a confirmed employment or for other work-related reasons (ii) demographic characteristics, namely gender and marital status, (iii) spatial factors including source-destination pairs and distance travelled, and (iv) migration experience as per years since migration. Our central hypothesis is that the internal migrants by being spatially flexible attempt to maximize the returns to their human capital endowments (particularly education) and minimize the penalty from being mismatched. But whether all kinds of migrants get similar results is primarily an empirical question and is yet to be explored. To the best of our knowledge, no one has investigated this aspect. To explore this dimension, the study uses data from India's National Sample Survey Office (NSSO) which was collected in 2007-08 and was primarily concerned with employment, unemployment, and migration particulars.

The contribution of this paper is twofold. Firstly, this paper contributes to the understanding of heterogeneity among migrants and the consequent differential impact of EOM in case of a developing country. Prior literature has argued that developing countries do not bias the migration patterns by providing unemployment benefits and other fiscal programs; and hence, serve as a better case to study the issues related to migration (Stark and Bloom, 1985). India with its marked inter-regional disparities provides a compelling case study for this analysis. Further, in India, it has been found that migration rate generally shows an increasing trend with the increase in education level, which indicates that higher educated workers are more likely to migrate (NSSO, 2010). However, the extent to which migrants are able to use their human capital, particularly education, is not well understood. Secondly, by analyzing the relationship between heterogeneity among migrants and the extent of and returns to EOM, we highlight the role of geographical limitations in affecting the opportunities to optimally gain the returns to attained education.[4]

There are only a few studies in India that capture the incidence of EOM. One such study is conducted by Bahl and Sharma (2021), where the authors analyze the extent of EOM in the Indian labor market and further analyze the role of EOM in understanding intra-education wage inequality (dispersion) using a quantile regression approach. They do not focus on the issue of spatial flexibility in the regional context and do not have migrants as their interest group. Our study differs from theirs along multiple dimensions. We focus on whether internal migration affects the incidence and returns to EOM in the context of India. The prime motivation behind this analysis is to test whether spatial flexibility leads to improvement in the returns to education in the wake of EOM. Further, as a corollary, we also analyze whether migration reduces the penalty for surplus education and improves the returns to required and attained education. Lastly, as discussed earlier, we estimate the heterogeneity in returns to education and depending on the various characteristics of internal migrants.

---

[4] It has been argued in the literature that by widening the labour market of a worker can improve the returns to education. Thus, in the past studies, it was hypothesized that migration will lead to better returns to education as this would enable the individuals to have access to a larger labour market. However, in this paper, we argue that not all migrants are the same and may still face geographical limitations which may lead to suboptimal returns to education.

Our study is an original contribution to the literature on EOM and spatial flexibility in the labor market. We contribute by highlighting the role of regional mobility and access to larger labor markets in reducing the mismatch and labor misallocation in the market while improving the pecuniary returns to education. Further, we emphasize that these returns are not uniformly accessible to all the migrants but are also affected by varying characteristics of the migrants themselves. This dimension needs to be understood better by the academicians and the policy makers for evidence-based targeted policymaking for the workers, especially migrants.

The rest of the paper is structured as follows. Section 2 describes the data source, measurement and empirical model. Section 3 explains the results. The final section presents the conclusion.

## 2. Data and Methods

This section describes the data source, some background on internal migration and EOM in India, followed by measurement of EOM, and empirical methodology.

### 2.1 Data source

In the case of India, the information regarding migrants and their labor market outcomes are available as part of the surveys conducted by the National Sample Survey Office (NSSO). The latest data collection period for this nationally representative survey is 2007-08 (64th round). The survey focuses on migration and its duration has been from July 2007 to June 2008 (64th round). Apart from this survey, till now, there is no other dataset that provides the micro data for the individual level analysis of the migrants and their comparison to the non-migrants. The other labor force surveys as well as the Census of India either do not capture the migration information or only provide the aggregated information at the district or state level.

The survey is conducted across 35 states and union territories of India and covers 125,578 households (79,091 in rural areas and 46,487 in urban areas), enumerating a total of 572,254 persons. The Census of India, 2001, has been used as a sampling frame for the survey. The NSSO survey contains detailed information on household characteristics such as religion, social group, land possessed, etc., along with the demographic characteristics such as education, age, gender, marital status, etc. The survey also provides information on the occupation and industry category of the workers.

Migration information is collected at two levels: household and individual. Considering the objective of this paper, the unit of analysis is at the individual level. The NSSO treats a member of the sample household as a migrant,

*"if he or she had stayed continuously for at least 6 months or more in a place (village/town) other than the village/town where he/she was enumerated. The village/town where the person had stayed continuously for at least 6 months or more prior to moving to the place of enumeration (village/town) was referred to as the "last usual place of residence" of that migrated person. Shifting of residence within village or/town was not considered as an event of migration"* (NSSO 2001, p. 14).

In other words, an individual is classified as a migrant if his or her last usual place of residence is different from the place of enumeration. The usual place of residence is a place where an individual has stayed continuously for at least six months.

For our analysis, the sample is restricted to the working-age population (15-59 years), which is consistent with the economically active age group as considered by the Government of India (NSSO, 2010). Further, we only consider workers employed for at least six months in the year preceding the survey, which has been measured by the principal activity status of the workers. Thus, we only consider the main workers (which is analogous to full time workers in an international context).

The main results consider the wage/salary of internal migrants who moved for work-related reasons[5] unless otherwise stated. The daily wages are reported for each of the 7 days preceding the survey, and we have calculated the average daily wage for the workers by summing their daily wages and dividing it by the number of days worked in the last 7 days. Given that the sample is only for the period of one year, we have measured the wages in nominal terms and have not used any Consumer Price Index deflator. Wages are measured in Indian rupees, which is a standard practice in the context of research on the Indian labor market. Additionally, while using the wages for estimation, we also drop extreme observations by truncating 0.5 percent at either end of the wage distributions. This ensures that our findings are not affected by the outliers on either side. We have not included self-employed workers in our analysis due to the non-availability of their earnings. We also provide estimates for non-migrants in the working-age group who receive wage/salary. Thus, our final sample size comprises 15,434 migrants and 60,689 non-migrants. Table A0 in the annexure contains the details about the sample considered in this study.

---

[5] We categorize a migration to be work-related if the reason to migrate is stated as any of the following – in search of a job, to take up a confirmed job, business, transfer of service/ contract, or proximity to place of work.

In addition to the NSSO survey, we also use the All-India Survey of Higher Education 2012 (AISHE) and Socioeconomic High-resolution Rural-Urban Geographic (SHRUG) database. These two sources provide us with instrumental variables for our empirical model. The variables used from the AISHE are the number of colleges/universities and autonomous institutions within a district. Given that our sample survey is for the period of 2007-08, we restrict the count of institutions till 2001, i.e., their year of establishment should be 2001 or before. The year 2001 is selected to get enough lag. The SHRUG database has provided the following instrumental variables: (i) calibrated night-time light in the year 2001 at the district level, and (ii) the standard deviation of night-time light across settlements within a district. A detailed discussion on these variables is provided in the sub-section on empirical model (sub-section 2.4).

## 2.2 Internal migration and education-occupation mismatch in India

In this subsection, we provide background about the phenomenon of internal migration and EOM in India. For the sake of brevity, we keep the discussion focused.

### Internal migration in India

As per the estimates from 2007-08 survey, around 26 percent (194 million) of rural and 35 percent (94 million) of urban population identify themselves as internal migrants in India. Around 55 percent of this migration is on account of marriage. Coming to the streams of migration, rural-rural is around 62 percent of total migration, followed by rural-urban (19 percent), urban-urban (13 percent), and urban-rural (6 percent) (Chandrasekhar and Sharma, 2015). In India, internal migration is largely an intra-state phenomenon. Intra-district migration is around 59 percent of total migration and inter-district (within state) has 29 percent share. Inter-state migration is only 12 percent of total migration.

In 2007-08, 28 million migrants migrated for work related reasons. Of these, around 91 percent were in the working age group 15-59 years, and around three-fourth considered urban areas as their destination.

Given these numbers, it can be said that internal migration is very large in India and is very heterogeneous in terms of destination choices and reason to migrate. For a more detailed discussion on internal migration, please refer to Chandrasekhar and Sharma (2015).

*Education-occupation mismatch in India*

In the context of developing countries like India, the misallocation of labor and subsequent mismatch between education of workers and occupational requirements remains a big concern. One extreme example of such instances is the application of around 3700 PhDs, 28,000 post-graduates and 50,000 under graduates for the posts of 62 messengers in the Uttar Pradesh (a state of India) police with minimum education requirement of class V (primary education- 5 years)[6]. Such instances have become a norm instead of being an anomaly. Based on the study conducted by Bahl and Sharma (2021) using the data from National Sample Survey Office's (NSSO) survey on employment and unemployment 2011-12, it has been estimated that around 19 percent (21 percent among men and 13 percent among women) of wage and salaried workers accounting for around 35 million were *over-educated* for the jobs. Additionally, the incidence of under-education for the same period was around 15 percent (26 million) of wage and salaried workers. These numbers are lower than many of the developed countries but remain quite high for the developing countries. The primary reason is the expansion in tertiary education along with the absence of job creation in the Indian economy.

In the next subsection, we describe the measurement of EOM in our survey data.

*2.3 Education-occupation mismatch: measurement*

This paper defines EOM as a gap between the attained years of education of an individual and the required years of education by her occupation. If attained education is greater than required education, a person is regarded as overeducated. If attained education is lower than the required education, the person is undereducated. A person is adequately educated when her attained and required education are aligned. Thus, to identify a worker's match status, we need two dimensions of crucial information, i.e., attained years of education and required years of education with respect to the occupation. NSSO collects information on the level of general education (no formal schooling, below primary, primary, middle, secondary, higher secondary, graduate, and postgraduate and above). We convert the level of education into years of formal education for our analysis.

For the measurement of required education, the literature provides three methods – workers' self-assessment (WA), job analysis (JA), and realized matches (RM) (see, Leuven and Oosterbeek, 2011 for complete discussion). WA captures the aspect of mismatch from workers' perspective, JA focuses on employers' side, and RM considers the perspective of a labor market. Thus, RM identifies the workers' match status by considering demand and supply side factors of a particular job. Given the focus of this paper, RM serves as the appropriate method due to the following reasons. First, the focus is on the nation-wide labor market outcomes of the migrants that can be best discussed using the measure which considers the entire labor market while estimating the required education. Second, in the case of developing countries, there is a significant dearth of surveys that capture the information required by WA and JA to measure the required education. Hence, RM remains the only applicable method. An added advantage is that the RM method is more common in the literature of spatial mobility and EOM (Nielsen, 2011; Poot and Stillman, 2016) and thus, would give us leverage to compare our results with the existing studies.

---

[6]    https://economictimes.indiatimes.com/news/politics-and-nation/over-93000-candidates-including-3700-phd-holders-apply-for-peon-job-in-up/articleshow/65604396.cms?from=mdr . Last accessed on Oct 31, 2021.

RM defines required education of an occupation as a range, where lower (upper) limit is calculated as the mean education attained by the workers minus (plus) one standard deviation in the education of workers working in a given occupation (Verdugo and Verdugo, 1989). Following this, the mean years of education for every three-digit National Classification of Occupation (NCO) 2004 codes have been computed using the sampling weights and a threshold of plus and minus one standard deviation from mean has been established to measure the range of required years of education. In this context, a worker will be categorized as adequately educated, if his or her attained education falls in the range of plus and minus one standard deviation from the mean, and a worker will be classified as overeducated (undereducated), if his or her attained education is higher (lower) than one standard deviation above (below) mean. The mean and standard deviation are measured only for the workers in the working-age group.

More precisely, suppose $e_i$ represents attained years of education of an individual 'i', $e_{nat}$ and $s_{nat}$ are the mean and standard deviation of years of education, respectively, for the occupation calculated at the national level. Thus, an individual will be considered:

Overeducated if: $e_i > e_{nat} + s_{nat}$;

Undereducated if: $e_i < e_{nat} - s_{nat}$; and

Adequately educated if: $e_{nat} - s_{nat} \leq e_i \leq e_{nat} + s_{nat}$

Using this definition and data from NSSO (2007-08)[7], we find that 71 percent of the wage/salary employed workers in India are adequately educated[8]. Further, despite having overall lower education levels as compared to other countries (Tilak, 2018), the problem of overeducation (16 percent) is more severe than under education (12 percent) in India. While analyzing EOM only for migrants, it is observed that migrants have higher incidence of education-occupation match as compared to the overall group. About 77 percent of migrants receiving wage or salary are adequately educated. However, for work-related migrants, the proportion of the adequately educated drops down to 70 percent. This highlights that individuals migrating for work-related reasons focus more on getting a job rather than an adequate one. This is indicated in the higher proportion of wage/salary employed (70 percent) among work-related migrants as compared to overall migrants (22 percent).

*2.4 Empirical Model*

*Ordinary Least Squares (OLS) Estimation*

This paper uses augmented Mincerian (Mincer, 1974) wage equation to estimate the returns to wage-determining characteristics. Thus, our wage equation is as follows:

$$log y_{ir} = \alpha + \beta_1 Edu_i + \beta_2 X_{1i} + \beta_2 X_{2r} + \beta_3 D_r + \beta_4 M_i + e_{ir} \quad (1)$$

where, the dependent variable is logarithm of daily wage of individual $i$ in region $r$. $Edu_i$ is our main interest variable, indicating years of education attained by an individual.

We have four sets of controls $X_{1i}, X_{2r}, D_r$ and $M_i$. The first set of controls are at the individual level, which includes years of formal education, age and its squared term[9], gender (male and female), marital status (unmarried, married, and others), interaction of gender and marital status, social group (scheduled tribe, scheduled caste, other backward class, and others), religion (Hindu, Muslim, Christian, and others), occupation categories at a broad level[10] (legislators, senior officials and managers, professionals, associate professionals, clerks, service workers and market

---

sales workers, skilled agricultural and fishery workers, craft and related trades workers, plant and machine operators and assemblers, and elementary occupations) and industry type (17 industry types including agriculture, fishing, manufacturing, and so on)[11]. The survey also provides information on employment status (employed, unemployed, and out-of-labour force) of migrants at their last usual place of residence. Since, being employed in the past can positively impact the wages, we include the dummy for being a wage/salaried employee at the last usual place of residence.

The second set of controls are at the regional level focusing on local labor market characteristics. This includes location sector (rural or urban), share of international immigrants[12] (measured at state level), log of working age population (at district level), and log of the labor force (at district level). These factors control for the heterogeneity in the labor market size, potential labor market competition as well as the interrelationship between international and internal migration.

The third set of controls use dummies for the fixed effect of origin and destination location. We use state level dummies for origin and destination locations.

The fourth set of controls are implemented in the empirical model for migrants to control for migrant heterogeneity in terms of reason for migration (to take a confirm job, search for work, and other reasons), migration distance[13] (intra-district, inter-district within state, inter-state), years since migration and migration stream (rural-rural, rural-urban, urban-rural and urban-urban).

Further, to capture the differential returns to EOM, Duncan and Hoffman (1981) suggested segregating the years of education into required years of education, surplus years of education (if a worker is overeducated), and deficit years of education (if a worker is undereducated), i.e.,

---

[11] The complete list is available at –
http://mospi.nic.in/sites/default/files/main_menu/national_industrial_classification/nic_2004_struc_detail.pdf
[12] This indicator is estimated at the state level to avoid the measurement error (attenuation bias) due to very low sample size at the district level. We are thankful to an anonymous reviewer for this suggestion.
[13] India is a federal country with three tiers of administrative bodies, which affect human mobility through various place-based policies, that is district, state, and central government. Based on this identification of boundaries, we can identify three types of migrants in our survey data. One, intra-district migrants who migrate within the boundaries of the district which is the smallest administrative unit observable in the data used. Second, inter-district (within state) migrants who cross the district boundaries but remain within the state in the process of migration. Third, inter-state migrants who cross the state boundaries. Based on these three definitions, generally, intra-district migrants travel the least distances while the inter-state migrants travel the most distances.

$$Edu_i^a = Edu_i^r + (0, Edu_i^s) - (0, Edu_i^d) \qquad (2)$$

where, $Edu_i^a$ and $Edu_i^r$ represent attained and required years of formal education, respectively, for an individual $i$. $Edu_i^s$ represents surplus years of formal education (which is measured as $Edu_i^a - Edu_i^r$) and $Edu_i^d$ represents the deficit years of formal education (which is measured as $Edu_i^r - Edu_i^a$).

Considering this, our wage equation becomes as follows:

$$logy_{ir} = \alpha + \delta_1 Edu_i^r + \delta_2 Edu_i^s + \delta_3 Edu_i^d + \beta_2 X_{1i} + \beta_2 X_{2r} + \beta_3 D_r + \beta_4 M_i + e_{ir} \qquad (3)$$

where, $Edu_i^r$ represents required years of formal education, $Edu_i^s$ represents surplus years of formal education, and $Edu_i^d$ represents deficit years of formal education. Other variables are interpreted as before. In equation (3), $\delta_1$, $\delta_2$, and $\delta_3$ depict the returns to required education, surplus years of education, and deficit years of education, respectively. Therefore, if a person is adequately matched, the years of surplus and deficit education would become zero. Further, if a person is overeducated, the years of attained education would be divided between required education and surplus education. Similarly, if a person is undereducated, the years of attained education would be divided between required education and deficit education. Thus, for every observation, the years of required, surplus, and deficit education would be largely uncorrelated[14].

The existing studies have usually found that (i) the returns to attained education are statistically lower than the returns to required education, (ii) although surplus education yield positive returns, the returns are statistically lower than that of the required education, and (iii) deficit education yields negative returns and this penalty is lower than return associated with required education (Hartog, 2000).

However, to estimate the unbiased coefficients in equation (3), there is a need of tackling a methodological issue of sample selection into employment (Heckman, 1979). This may arise as the wage of a worker is observable only after getting employed. But there is a possibility that some unobserved factors (such as ability) are positively related to both wages and likelihood of being employed. Therefore, considering only the employed individuals for analysis would lead to biased estimates. Another problem in our data is the unavailability of income information for self-employed individuals. However, the choice of self-employment versus wage/salary employment is not random and disregarding the possible sample under self-employment may again lead to sample selection bias (Dolton and Makepeace, 1990). Further, the decision to migrate is also not random. Hence, there is a problem of triple-selection bias. To overcome this problem, Heckman correction procedure is applied for all the three decisions, i.e., the decision to work (to be or not to be engaged in economic activity), the choice of economic activity status (wage/salary or self-employment), and the decision to migrate (to be or not to be migrant).

---

[14] Interestingly, in the literature on EOM, studies have never discussed the concern of potential correlation between required, surplus, and deficit years of education. We are thankful to an anonymous reviewer for raising this concern. In our dataset, we have observed that the pairwise correlation of required years of education with surplus and deficit years of education is -0.08 and -0.09, respectively, which is small. However, the pairwise correlation between surplus and deficit years of education is 0.5. This will increase the standard error of the estimated coefficient by the variation inflation factor (VIF = 1(1-r2), where r is a pairwise correlation) of 1.33. Thus, the concern of multicollinearity, in this case, is not present.

To use this method, it is a prerequisite to identify at least one variable that does not affect the wages but influences the probability of participation (in our case, employment, self-employment, and migration); such a variable is called an exclusion variable(s). We have used the number of dependent members (under 15 and over 59) in a household, household type (self-employed, regular wage/salary earning, casual labor, and others), and household size as exclusion variables for the decision to work. This is in line with the literature that advocates the use of family characteristics as appropriate exclusion variables for the choice of work (Buchinsky, 2002). Further, we have used land possessed by the household as an exclusion variable for choice of economic activity status. This is on the grounds that land ownership improves the access to credit (Feder and Onchan, 1987), as it can be used as a collateral (Kaas et al., 2016). Further, self-employment being riskier than wage/salary employment, having access to a land can be considered as a safety cushion. We have used the migrant network indicators as the exogenous predictors for the decision to migrate (Bertoli, 2010; Bertoli et al., 2013; McKenzie and Rapoport, 2010). For the discussion on measurement of migrant network indicators, please refer to section A2 of the annexure. In short, to address the self-selection, we have estimated the first stage sample selection for the Heckman correction using the exclusion variables and then the resulting correction terms are used in the final wage equation.

Therefore, our final wage equation is as follows:

$$logy_{ir} = \alpha + \delta_1 Edu_i^r + \delta_2 Edu_i^s + \delta_3 Edu_i^d + \beta_2 X_{1i} + \beta_2 X_{2r} + \beta_3 D_r + \beta_4 M_i + \lambda_i^{emp} + \lambda_i^{wage} + \lambda_i^{mig} + e_{ir} \quad (4)$$

where, $\lambda_i^{emp}, \lambda_i^{wage}$ and $\lambda_i^{mig}$ are the correction terms for sample selection related to employment (versus not employed), wage/salary employment (versus self-employment), and migration (versus no migration), respectively. Other variables are interpreted as before. The estimates of the sample selection probit models for all three decisions are provided in the annexure (see Table A2).

It can be noticed that the exclusion variables for all the three decisions are statistically significant. Furthermore, apart from the exclusion variables, we have also controlled for years of education, age, age squared, gender, marital status, interaction of gender and marital status, social group, religion, state, and sector.[15] All these variables appear to be statistically significant in influencing the probability of migration, employment, and wage/salary employment.

The equation (4) is estimated separately for a varied group of migrants. The summary statistics for the wage/salary employed non-migrants and work-related migrants in the working-age group (15-59) are presented in Table A3 of annexure, for the sake of brevity. Also, the incidence and average daily wages (in Indian rupees) of work-related migrants in the working age-group by different demographics are given in Table A4 of annexure.

*Instrumental variable (IV) estimation*

In our ordinary least square (OLS) model, after accounting for the sample selection bias regarding decision to employment, nature of employment (wage/salary versus self-employment) and migration, the concern for endogeneity of the main variable, i.e., years of education persists. In the empirical literature on returns to education, instrumental variable estimation is the best way to address this concern. To correct the endogeneity of general education measured as the attained years of education, we require exogenous variables as instruments in our model. To prove their validity and strength, these instruments should meet the following conditions. (1) They should be uncorrelated to the error term, also called the orthogonality condition. (2) They must be correlated to the endogenous variable that needs to be instrumented. (3) They should not be part of the original model, also referred to as the exclusion criteria. (4) The number of instruments should be at least equal or more than the number of endogenous variables. The finding of instruments in the observational data that simultaneously satisfy the above-mentioned four conditions remains a major problem in implementing instrumental variable models for identification.

For our instrumental variable estimation, we have identified the following instrument variables based on the literature: (i) geographic variation in the number of colleges/educational institutions

---

at the district level, (ii) monthly per capita expenditure as a proxy for household status, (iii) calibrated night-time light in the year 2001 at the district level, and (iv) standard deviation of night-time light across the settlements within a district.

However, one additional statistical concern for our instrument variable model is weak instrument condition. To overcome this, we have taken an approach suggested by Lewbel (2012). From hereafter, we will refer to the instrumental variable model as the Lewbel IV model.

Further, we have faced one additional problem while estimating the Lewbel IV model for the migrants. For the non-migrants, the district of schooling and work remains the same, and therefore, the number of colleges/educational institutions and night-time light can be used as the valid instrumental variables. However, the same is not true for all the migrants, especially inter-district and inter-state migrants. Also, the survey does not provide us with an identifier of their origin district.[16] Thus, using the above instrumental variables for inter-district and inter-state migrants will cause measurement error in the instrument variables leading to biased and inconsistent Lewbel IV estimates. Given this caveat, the instrumental variables used in our model remain only valid for the non-migrants and the intra-district migrants. Hence, in our results section, we have only reported the results for these two groups.

The detailed equations for Lewbel IV models and corresponding discussion are provided in Section A3 of Annexure.

In the next section, we present the results of sample selection models, followed by the OLS models, and finally Lewbel IV estimates along with the diagnostic tests for the instrumental variable models.

## 3. Results

In this section, we present the wage returns to education and EOM for non-migrants and work-related migrants segregated by their demographic characteristics, spatial factors, and migration experience taking account of other wage-determining characteristics. For the sake of brevity, the estimates for only concerned variables are shown.[17]

---

[16] In the study by Card (1993), they had past information about the geographic location of the household, which was used to identify the corresponding distribution of schools for the households. However, such information is not available in our survey.

[17] The results for the full set of controls are available in the annexure (Table A5-A10).

Before going ahead with the main results, the sample selection models (see Table A2 in annexure) are briefly discussed. While selecting the model for migration, we have found that the migrant network has an inverted-U shaped relationship with the decision to migrate (Yamauchi and Tanabe, 2008). While analyzing the employment selection model, we have found that individuals with a higher number of dependents in the household have a higher probability of entering the labor market. Further, individuals with higher land endowments are more likely to engage in wage and salary employment (Table A2, column 3). This can be explained by the livelihood diversification strategy (Reardon, 1997). We have included the inverse mills ratio from these three sample selection models in the later stage of econometric analysis and found that mills ratio for migration and employment are significant with negative and positive coefficients, respectively. This indicates that there is no significant sample selection bias depending on the type of employment. However, migration has negative sample selection bias and employment decision has positive sample selection bias, which has been corrected through this Heckman procedure.

*3.1 Ordinary Least Square (OLS) estimates*

Having discussed these sample selection models, the main results are reviewed. Firstly, we present the results for work-related migrants and non-migrants (Table 1). It is found that on an average, for every one year of formal education, the work-related migrants earn a wage reward of three percent, while for non-migrants, the return is around two percent. Further, dividing the years of education into required, surplus, and deficit years of education, it is observed that the results for both migrants and non-migrants are aligned with the literature (Hartog, 2000; Leuven and Oosterbeek, 2011). The returns to required education are higher than the returns to attained education. Also, while the returns to surplus education are positive, they are lower than the returns to required education. The returns to deficit education are negative, which highlight that the undereducated workers get lower wages as compared to adequately and overeducated workers in a given occupation. In the subsequent analysis, the migrants have been segregated into various groups to find answer to the primary question of this paper.

**<Table 1 Here>**

First, we differentiate the migrants according to the reason to migrate (Table 2), which are (i) in search of a job (Table 2 - Column 2); (ii) to take-up a confirmed job (Table 2 - Column 3); and (iii) other job-related reasons (Table 2 - Column 3) that includes business, transfer of job, and proximity to workplace[18]. Migrants who did not have a confirmed job before migration earn lower returns to education as compared to migrants who had one. Migrants in search of a job have lower bargaining power in the labor market (Blanchard, 1991) due to their spatial inflexibility and urgency to get a job for covering their living expenses at the destination. Further, migrants with a confirmed job get higher returns with respect to required and surplus education and higher penalty for deficit education. Migrants for other reasons (business, transfer, proximity to work) have higher returns than the other two categories for attained education, but get similar returns for required and surplus years of education. While the difference in the returns to deficit education is significant, all the groups of migrants earn statistically similar returns to their required and surplus education.

**<Table 2 Here>**

For spatial factors, the migrants by source-destination pairs, that is, rural-rural, rural-urban, urban-rural, and urban-urban are separated (Table 3). The rural and urban areas vary in terms of job opportunities, infrastructure and so on. This necessitates the consideration of source-destination pairs while analyzing the returns to education. We have found that irrespective of the destination, workers from urban areas earn higher returns to education. This could be due to better quality of education in the urban areas as compared to the rural areas (Agrawal, 2014). Also, workers in urban areas are better informed regarding the availability of job opportunities, and hence, can choose the more appropriate alternative. Besides, workers who move from urban areas to urban areas witness the highest penalty for deficit years of education. We have also found differences in the returns to be significant across all the groups. Further, workers moving to rural areas experience a more prominent difference between their returns for required and surplus education as compared to other groups. One plausible explanation is that jobs in rural areas are not as skill-intensive as jobs in urban areas; and consequently, wages are determined more by the job characteristics rather than the education of workers. Thus, having surplus education does not add much returns in terms of wages.

**<Table 3 Here>**

---

[18] In section A4 of annexure, we also provide the estimates for returns to education for other than work-related migrants, which include education, forced migration (due to natural disaster, socio-political problems and displacement for development projects), marriage, tied movers (due to mobility of parents/family members) and others.

Further, to estimate the differences in the returns as per the distance travelled in migration, we have divided the migrants into three groups (Table 4): intra-district (column 2), inter-district but within state (column 3), and inter-state (column 4).[19] It is found that workers who move inter-district but within the state earn the highest return on education. This phenomenon can be explained using the contrasting results found in case of the international and internal migrants (Devillanova, 2013). Workers who choose to migrate only within their district have lower intensity of spatial flexibility, and thus get lower returns to education as compared to workers who migrate inter-district. But as distance increases, the negative impact of imperfect portability of human capital endowments begins to dominate the positive impact of spatial flexibility on wages. Workers moving to a different state may find it difficult to transfer their human capital due to variation in the quality of education, lack of language proficiency, cultural differences, and so on (Krishna, 2004); and therefore, earn lower returns. Further, since education of the inter-state migrants is not directly transferable, they also earn a lower penalty for deficit years of education. However, we have found that differences in the returns to surplus and deficit education are not significant among these groups. Therefore, it can be concluded that workers who are spatially flexible and have better portability of their human capital endowments earn the highest rewards for the required education.

**<Table 4 Here>**

Next, conditional on age, we have found that the longer a migrant has been at a particular destination, the higher are the returns to education (Table 5). This is in line with the literature that claims that migrants' wages improve with the time spent at destination (Borjas et al., 1992; Yamauchi, 2004) on account of the assimilation process. The researchers have also found that the difference in the likelihood of being mismatched between migrants and non-migrants decreases with the increase in the stay of migrants (Aleksynska and Tritah, 2013; Nieto et al., 2015). We have also observed that the returns to required and surplus education for the group are higher with respect to the highest migration experience. However, the differences in the returns are insignificant across groups.

**<Table 5 Here>**

---

[19] The district is the smallest geographical unit at a sub-national level. In 2007–2008, India comprised 35 states and union territories, 87 NSS regions and 588 districts.

Therefore, to summarize, while migrants get differential returns to attained, required, surplus, and deficit education on account of their respective reason to migrate and spatial factors, the migration experience does not lead to differential returns.[20]

3.2 Lewbel IV estimates

3.2.1 *Test for endogeneity*

Before going ahead with the 2SLS IV model, the endogeneity of the variable - attained years of education for migrants and non-migrants has been tested using the Durbin-Wu-Hausman test of endogeneity (Cameron and Trivedi, 2010). For the non-migrants, we have observed that the attained years of education is endogenous, as the null hypothesis that the variable is exogenous is rejected. The Durbin-Wu-Hausman test of endogeneity F-statistic (p-value in parenthesis) values are 8.26 (0.004),. For the migrants, the F-statistics value is 14.75 (0.0001), which indicates that attained years of education is endogenous. Additionally, for the intra-district migrants also, we find that years of education is endogenous (4.0, p-value 0.046).

Next, we have tested the endogeneity of variables: required years of education, surplus years of education and deficit years of education. Among the non-migrants, we have failed to reject the null hypothesis for deficit and surplus years of education, which means that these variables are exogenous but the required years of education is endogenous. The F-statistic (p-values in parenthesis) values of required, surplus and deficit years of education for non-migrants are 1.87 (0.10), 0.03 (0.85) and 0.3 (0.82), respectively; For the migrants, the F-statistic values for required, surplus and deficit years of education is 2.04 (0.15), 2.26 (0.13) and 1.58 (0.11) respectively. This indicates that these three variables are exogenous in our sample. For the intra-district migrants, F-statistic values for required, surplus and deficit years of education is 0.14 (0.71), 2.6 (0.10) and 4.4 (0.06) respectively. Thus, for the smaller sample of intra-migrants also, we observe that these three variables are exogenous.

The endogeneity tests indicate that instrument variable model is required for the attained years of education. However, for the required, surplus and deficit years of education, an instrument variable model is not needed due to the exogeneity of these variables in our model.

---

[20] We have also analyed the returns to education and returns to EOM for married migrants and by gender. The results are discussed in the section A5 and A6, respectively, in the annexure.

*Results from Lewbel IV estimates*

It is clear from the endogeneity test that only attained years of education is endogenous. Further, due to measurement error issues, we have only run the Lewbel IV regression for non-migrants and intra-district migrants.

With respect to the instrumental variable estimation for attained years of education for migrants and non-migrants, it is observed that OLS estimates are different from Lewbel IV estimates (see Table 6). For the non-migrants, Lewbel IV estimates are marginally smaller than OLS estimates. For the migrant workers, the Lewbel IV model only for the intra-district migrants has been estimated because there is no information about the origin district of in-migrants so, use of instrument variables for these migrants will be incorrect. We know the district of origin and destination for the intra-district migrants only by definition. Therefore, we can use the instrument variables for this sub-sample of migrants. We observe that Lewbel IV estimates for intra-district migrants are higher than OLS estimates, highlighting that OLS estimates were downward biased (see Table 6). The yearly returns to attained education for the intra-district migrants is around 4 percent as compared to OLS estimates of around 3 percent.

**<Table 6 Here>**

Ideally, we would want the instrument variable estimation for the full sample available to us, i.e., for both non-migrants and migrants. However, information about the district of origin is not available for all the migrants, we are only aware of their state of origin, except for intra-district migrants. Further, aggregating the instrumental variables at the state level would make them correlated to other state level variables, thus violating the exogeneity condition. In addition, it would reduce the variation in the instrument variables, thereby reducing their predictive power leading to large standard errors and imprecise/biased estimates. Moreover, if we consider the current district's information for the IV estimation of migrants, it will lead to measurement error because the place of residence may not be the same as the place of education for them. Given this caveat, the instrument variable estimation is only used for the intra-district migrants and non-migrants, whose origin district is the same as the destination district, and therefore, we can be sure about the validity of the instrument variables.

Lastly, we can draw two conclusions from our instrument variable model. First, for the non-migrants, the Lewbel IV estimates are marginally lower than OLS estimates. For the migrants, attained years of education is endogenous and the Lewbel IV estimates for migrants are higher than OLS. This indicates that the estimates for the attained years of education for migrants are higher than non-migrants, and there are higher returns to being spatially flexible. In other words, the intra-district migration increases the returns to human capital and leads to higher productivity gains for the labor markets. Second, we have found that EOM indicators, i.e., required, surplus and deficit years of education are exogenous in our sample, and therefore, OLS estimates are unbiased for these variables. This means that the majority of our results remain unchanged and the core results of our study are robust.

## 4. Conclusion

This paper analyzes the extent to which migrants can utilize their education, especially in the context of their occupation. The utilization is tested using the EOM framework. In particular, we have investigated the incidence of EOM and returns to education and EOM for migrants considering the heterogeneity among them.

The main results are as follows. First, while the incidence of EOM (undereducation and overeducation) does not differ much between migrants and non-migrants, migrants with different reasons to migrate, demographic characteristics, spatial factors and migration experience witness markedly different rates of EOM. Second, migrants with a confirmed job at the destination before migration earn higher returns to attained, required, and surplus education as compared to migrants who first migrate and then search for a job. Third, workers (females and higher educated groups) who get higher returns to required and surplus education also get higher penalties for their deficit years of education. Fourth, we have unveiled that irrespective of the destination, workers from urban areas earn higher returns to education, but they are the ones who pay the highest penalty for having deficit education. Further, while the labor market penalizes or rewards EOM equally in terms of statistics irrespective of the distance travelled, the returns to attained and required education vary significantly. Workers who choose to migrate only intra-district (relatively spatially inflexible) get lower returns to education as compared to workers who migrate inter-district (spatially flexible). But as the distance increases, the negative impact of imperfect portability of human capital endowments begins to dominate the positive impact of spatial flexibility on wages. Next, migration experience and assimilation does not lead to differential returns to education and EOM. We have also used the Lewbel IV model to correct the endogeneity of the attained years of education for the intra-district migrant and non-migrants in our sample. The Lewbel IV estimates are marginally higher for the intra-district migrants, whereas lower for the non-migrant as compared to the OLS estimates. Lastly, we have found that the required, surplus and deficit years of education are exogenous in our sample and therefore, do not require instrumental variable estimation.

The key implications of this study are as follows. First, by highlighting the differential in the labor market outcomes for the varied migrants, the study stresses the need that while analyzing the decision to relocate, individuals should consider these differences in the returns to attain the maximum benefits. For example, it would be beneficial for the individuals to have a confirmed job at a destination before they decide to migrate. Also, to achieve the optimal gains to education, an individual should move to a place where he or she will be able to adequately transfer the human capital. Second, while spatial flexibility is seen as an important policy tool to redistribute the human resources for achieving higher productivity and growth for a national economy as a whole (Maxwell, 1988; Sahota, 1968), the under-utilization of human capital endowments of migrants could reverse such effects. The literature has found that apart from lower wages, EOM can lead to various other adverse consequences such as lower productivity (Tsang, 1987), higher attrition (Verhaest and Omey, 2006), etc. Hence, adequate attention has to be paid on the migrants' EOM to achieve the desired results.

The generalizability of results has always been a concern for country specific studies, that is, whether the results have external validity. With respect to this, two sets of arguments can be provided. First, the results are not specific to the nature of internal migration being measured. Similar results are also obtained by a number of studies as cited in the results section (Aleksynska and Tritah, 2013; Devillanova, 2013; Hensen et al., 2009; Nieto et al., 2015). Further, we have explored a number of dimensions of internal migrants based on their characteristics (reason for migration, migration distance, years since migration and migration stream). Thus, we provide more comprehensive analysis of internal migrants and EOM. All the characteristics observed or measured in the migration surveys can be adopted with respect to country-specific characteristics. Thus, the findings from India may be useful depending on the nature of internal migration. Additionally, in our study, we have controlled for a number of regional and local labor market, individual, household, migration specific variables, thereby providing a robust set of results. Thus, our study will surely be useful to researchers from other countries, mainly developing countries, who would be doing research on internal migration. Second, it can be argued that India accounts for around 17 percent of the world population and is the second-largest country by population. Further, out of the estimated 740 million internal migrants worldwide in 2009 (Klugman, 2009: Human Development Report), around 280 million (estimates based on NSSO (2010), as per the 2007-08 survey happen in India. This is slightly more than one-third of the total internal migrants in the world. Hence, our data is representative of these internal migrants. Therefore, understanding their labor market outcomes will provide insights for studies on internal migration especially in developing countries.

Although the study aims to provide a rich description of internal migration and EOM, the limitations are inevitable. The lack of information on skills and quality of education makes it difficult to differentiate migrants with similar levels of education. Therefore, it may be possible that overeducated workers are under-skilled, which makes them earn lower wages. However, this dimension cannot be explored due to the lack of data on the skill level or quality of education. Another important aspect is to account for other sources of human capital formation apart from formal education in these models, especially for the developing countries, where informal training and on the job learning remains important in determining the match between workers and their jobs[21]. However, converting these various dimensions into a single index value for the purpose of comparison is a tedious task. Lastly, there is an unavailability of information on spatial aspects of the labor market. For example, having geocoded labor market datasets for such analysis would help in understanding the spatial dependence across various local labor markets, and their implications for misallocation of labor and corresponding returns to human capital. Information on the spatial location of the migrants at the destination can help us understand the mechanisms through which EOM are corrected or accentuated. Further, the probability of employment and consequent match status is affected by the prevailing conditions in the local labor market. Therefore, the spatial information on labor market aspects can further enrich this study. Lastly, in the context of instrumental variable models, we can only use the non-migrant and intra-district migrants' sample for our analysis due to lack of information about the district of origin of the migrant individuals. This has restricted our analysis for Lewbel IV estimate to the sub-sample and we cannot estimate models for heterogeneity along other dimensions in our study.

---

[21] We are thankful to an anonymous reviewer for highlighting this important aspect of labor markets in the developing countries.

Table 1: Returns to education for wage/salary employed in working-age group: Work-related migrants and non-migrants

| Explanatory variables | Dependent variable: ln(wage) | |
| --- | --- | --- |
| | (1) | (2) |
| | Work related migrants | Non-migrants |
| Attained education | 0.0298*** | 0.0244*** |
| | (0.00182) | (0.000937) |
| Observations | 13,887 | 56,440 |
| R-squared | 0.501 | 0.359 |
| Required education | 0.0751*** | 0.0764*** |
| | (0.00584) | (0.00477) |
| Surplus education | 0.0324*** | 0.0227*** |
| | (0.00334) | (0.00194) |

| Deficit education | -0.0241*** | -0.0238*** |
|---|---|---|
|  | (0.00357) | (0.00194) |
| Observations | 13,745 | 53,824 |
| R-squared | 0.512 | 0.375 |

Table 2: Returns to education and EOM for employed migrants in working-age group: by reason to migrate (work related)

| Explanatory variables | Dependent variable: ln(wage) | | | | |
|---|---|---|---|---|---|
|  | (1) | (2) | (3) | (4) |  |
|  | Overall | Job Search | Confirm Job | Others | Chow test |
| Attained education | 0.0298*** | 0.0204*** | 0.0337*** | 0.0433*** | 21.34*** |
|  | (0.00182) | (0.00230) | (0.00325) | (0.00561) |  |
| Observations | 13,887 | 6,635 | 4,104 | 3,148 |  |
| R-squared | 0.501 | 0.369 | 0.482 | 0.377 |  |
| Required education | 0.0751*** | 0.0648*** | 0.0787*** | 0.0770*** | 1.22 |
|  | (0.00584) | (0.0104) | (0.00918) | (0.00913) |  |
| Surplus education | 0.0324*** | 0.0255*** | 0.0304*** | 0.0375*** | 1.99 |
| | (0.00334) | (0.00462) | (0.00619) | (0.00744) | |
| Deficit education | -0.0241*** | -0.0153*** | -0.0314*** | -0.0463*** | 7.76*** |
| | (0.00357) | (0.00389) | (0.00604) | (0.0163) | |
| Observations | 13,745 | 6,541 | 4,071 | 3,133 | |
| R-squared | 0.512 | 0.384 | 0.490 | 0.388 | |

Source: Authors' calculation based on NSSO employment, unemployment and migration survey, 2007-08.
Note:
 (i) *** signals significant at 1% level ,** signals significant at 5% level and * signals significant at 10% level.
 (ii) Robust standard errors are given in parenthesis.
 (iii) Chow test indicates whether the difference in the coefficients is significant or not.

| | (1) Overall | (2) Rural-rural | (3) Rural-urban | (4) Urban-rural | (5) Urban-urban | Chow test |
|---|---|---|---|---|---|---|
| **Table 3: Returns to education and EOM for wage/salary employed migrants in working-age group: by migration stream** | | | | | | |
| **Dependent variable: ln(wage)** | | | | | | |
| Explanatory variables | | | | | | |
| Attained education | 0.0298*** | 0.0242*** | 0.0241*** | 0.0305*** | 0.0460*** | 19.01*** |
| | (0.00182) | (0.00307) | (0.00261) | (0.00639) | (0.00478) | |
| Observations | 13,887 | 3,937 | 5,715 | 1,086 | 3,149 | |
| R-squared | 0.501 | 0.498 | 0.424 | 0.587 | 0.505 | |
| Required education | 0.0751*** | 0.0699*** | 0.0725*** | 0.0936*** | 0.0750*** | 1.53 |
| | (0.00584) | (0.0115) | (0.00915) | (0.0170) | (0.0110) | |
| Surplus education | 0.0324*** | 0.0222*** | 0.0332*** | 0.0233** | 0.0486*** | 8.09*** |
| | (0.00334) | (0.00645) | (0.00505) | (0.0108) | (0.00762) | |
| Deficit education | -0.0241*** | -0.0230*** | -0.0140*** | -0.0322** | -0.0402*** | 6.04* |
| | (0.00357) | (0.00647) | (0.00384) | (0.0162) | (0.0121) | |
| Observations | 13,745 | 3,865 | 5,680 | 1,064 | 3,136 | |
| R-squared | 0.512 | 0.501 | 0.441 | 0.601 | 0.517 | |

Source: Authors' calculation based on NSSO employment, unemployment and migration survey, 2007-08.
Note:
 (i) *** signals significant at 1% level, ** signals significant at 5% level and * signals significant at 10% level.
(ii) Robust standard errors are given in parenthesis.
(iii) Chow test indicates whether the difference in the coefficients is significant or not.

Table 4: Returns to education and EOM for wage/salary employed migrants in working-age group: by distance

| | | | Dependent variable: ln(wage) | | |
|---|---|---|---|---|---|
| | (1) | (2) | (3) | (4) | |
| Explanatory variables | Overall | Intra-district | Inter-district (within state) | Inter-state | Chow test |
| Attained education | 0.0298*** | 0.0282*** | 0.0340*** | 0.0270*** | 3.09 |
| | (0.00182) | (0.00330) | (0.00324) | (0.00268) | |
| Observations | 13,887 | 4,592 | 4,986 | 4,309 | |
| R-squared | 0.501 | 0.530 | 0.530 | 0.451 | |
| Required education | 0.0751*** | 0.0874*** | 0.0821*** | 0.0531*** | 8.96*** |
| | (0.00584) | (0.0100) | (0.00938) | (0.00850) | |
| Surplus education | 0.0324*** | 0.0361*** | 0.0291*** | 0.0325*** | 0.77 |
| | (0.00334) | (0.00546) | (0.00593) | (0.00553) | |
| Deficit education | -0.0241*** | -0.0166** | -0.0338*** | -0.0212*** | 4.24 |
| | (0.00357) | (0.00690) | (0.00621) | (0.00489) | |
| Observations | 13,745 | 4,518 | 4,953 | 4,274 | |
| R-squared | 0.512 | 0.540 | 0.541 | 0.464 | |

Source: Authors' calculation based on NSSO employment, unemployment and migration survey, 2007-08.
Note:
(i) *** signals significant at 1% level, ** signals significant at 5% level and * signals significant at 10% level.
(ii) Robust standard errors are given in parenthesis.
(iii) Chow test indicates whether the difference in the coefficients is significant or not.

Table 5: Returns to education and EOM for wage/salary employed migrants in working-age group: by years since migration

| Explanatory variables | (1) Overall | (2) 0-2 | (3) 3-5 | (4) 6-10 | (5) 11 and above | Chow test |
|---|---|---|---|---|---|---|
| | | | Dependent variable: ln(wage) | | | |
| Attained education | 0.0298*** | 0.0303*** | 0.0276*** | 0.0240*** | 0.0315*** | 2.8 |
| | (0.00182) | (0.00310) | (0.00375) | (0.00369) | (0.00335) | |
| Observations | 13,887 | 4,714 | 2,875 | 2,553 | 3,745 | |
| R-squared | 0.501 | 0.506 | 0.532 | 0.548 | 0.506 | |
| Required education | 0.0751*** | 0.0699*** | 0.0725*** | 0.0936*** | 0.0750*** | 5.4 |
| | (0.00584) | (0.0115) | (0.00915) | (0.0170) | (0.0110) | |
| Surplus education | 0.0324*** | 0.0222*** | 0.0332*** | 0.0233** | 0.0486*** | 5.41 |
| | (0.00334) | (0.00645) | (0.00505) | (0.0108) | (0.00762) | |
| Deficit education | -0.0241*** | -0.0230*** | -0.0140*** | -0.0322** | -0.0402*** | 3.86 |
| | (0.00357) | (0.00647) | (0.00384) | (0.0162) | (0.0121) | |
| Observations | 13,745 | 3,865 | 5,680 | 1,064 | 3,136 | |
| R-squared | 0.512 | 0.501 | 0.441 | 0.601 | 0.517 | |

Source: Authors' calculation based on NSSO employment, unemployment and migration survey, 2007-08.
Note:
(i) *** signals significant at 1% level, ** signals significant at 5% level and * signals significant at 10% level.
(ii) Robust standard errors are given in parenthesis.
(iii) Chow test indicates whether the difference in the coefficients is significant or not.

Table 6: Returns to education (Lewbel IV Model): Work-related Intra district migrants and non-migrants

| | Dependent variable: ln(wage) | | | |
|---|---|---|---|---|
| | (1) | | (2) | |
| Explanatory variables | Non-migrants | | Intra-district work related migrants | |
| | OLS | OLS | OLS | Lewbel IV |
| Attained education | 0.0244*** | 0.0219*** | 0.0291*** | 0.0396*** |
| | (0.000937) | (0.00405) | (0.0032) | (0.00777) |
| Observations | 56,440 | 4,592 | 4,592 | 54,320 |
| R-squared | 0.359 | 0.528 | 0.528 | 0.351 |

Source: Authors' calculation based on NSSO employment, unemployment and migration survey, 2007-08.
Note: (i) *** signals significant at 1% level and ** signals significant at 5% level
(ii) We only report the second stage estimates for the interest variable. The detailed estimates are provided in the annexure table A10.

**Annexure A**

**Section A1**

**Sensitivity check: Effect of changing the threshold for measurement of required education for an occupation**

We have measured the sensitivity of EOM by using different thresholds over and below required years of education (measured by mean years of education). Please refer to Table A11 for detailed numbers.

Using the threshold of one standard deviation over and below the mean years of education for the occupation categories, it is found that among the migrants, the match (adequate) was around 70 percent and the incidence of over and under educated individuals is 15 percent each.

A change of threshold to 0.9 standard deviation reduces the adequately educated (match) to 63 percent for migrants. Also, the under and overeducated proportion increases to 17 and 20 percent, respectively.

Now, changing the threshold to 1.1 standard deviation increases the incidence of adequately educated (match) to 78 percent, the undereducated to 11 percent, and the overeducated to 11 percent of the migrants.

Lastly, using one standard deviation from median, we have found that the incidence of adequately educated (match) is 67 percent of migrants, the undereducated are 19 percent, and the overeducated are 14 percent.

**Section A2**

**Measurement of migrant networks**

For the migrant network indicator, we have used the stock of migrants at the destination weighted by their experience at the destination. Experience is measured as years since migration.

This indicator is widely used in the literature for measurement of the informational effect of the migrant networks. There are two channels through which migrant networks affect the decision to migrate. First, individuals use personal contacts and social ties in the process of migration. The existence of a large network at the destination eases that process and reduces the cost of mobility. Second, a large migrant size also acts as a signal to the potential migrant that the destination is attractive and this induces 'herd effect' as termed in this literature (Calvo-Armengol and Zenou, 2005; Iversen et al., 2009; Yamauchi and Tanabe, 2008). But the migrant network can also exert a negative effect on the decision to migrate. The past studies such as Wahba and Zenou (2005), Yamauchi and Tanabe (2008) also show that if the network size is too large (measured by stock of migrants), it may also lead to negative impact indicating the congestion effect and competition for employment at the destination among the migrants with similar skill sets.

Therefore, in the literature, the effect of network size is considered to be inverted U shape. To capture that effect, we have used migrant stock weighted by the experience and the squared term of that variable in our sample selection model (Sharma and Das, 2018; Yamauchi and Tanabe, 2008).

$$Migrant\ network_r = \Sigma_s^N \Sigma_i^J Exp_i Migrant_{isr}$$

In this equation, $i$ denotes individual migrant, $s$ denotes source region (there are N such regions - states), $r$ denotes destination region (there are J such regions -district). $Exp$ denotes years since migration for a particular migrant. $Migrant$ indicates the status of the individual being migrant or not.

**Section A3**

**Additional discussion on Instrumental Variable (IV) estimation**

*Instrument Variables*

In our study, for the exogenous instrument variables, two data sources have been relied on, which can be combined with the existing survey data for the identification strategy. The first source of information is the All-India Survey of Higher Education 2012 (AISHE) database. This database provides the geographic distribution of institutions of higher education, i.e., colleges, universities, and autonomous/independent institutions within the country. Using this database, the district wise distributions of colleges and institutions can be identified and combined with the migration survey. Drawing insights from the study of Card (1993)[22], we have implemented the geographic variation in the access to colleges/educational institutions as an instrumental variable for access and years of education to exogenously measure the attained education of the individuals. Apart from these, as suggested by Card (1993), only access to schools is not sufficient to ensure an increase in years of schooling for the residents, but the status of the household also matters. A richer household with access to schools and colleges will gain more in terms of increased years of schooling than a poorer household. Therefore, we have also used the proxy for household income status measured by monthly per capita expenditure as an instrumental variable. In addition, the interaction term between household income status and number of colleges/educational institutions in the district of residence has also been applied in the analysis.

Further, there is need of additional instrument variables for the required, surplus, and deficit years of education variables, which is nothing but the three components of attained years of education. With increase in the years of education for individuals, their probability of being overeducated (surplus years of education) for the jobs monotonically increases, while the inverse is true for undereducation (deficit years of education). Based on the condition that the number of instruments should be at least equal or more than the number of endogenous variables, we require at least three valid and strong instrument variables for the exact or over-identification of the IV model. To fulfil this condition, additional instrument variables have been searched, which has taken us to the second data source. This data source is the Socioeconomic High-resolution Rural-Urban Geographic (SHRUG) database that provides the calibrated night-time light data at the district and sub-district levels for India (Asher et al., 2021). We have used two indicators of night-time light as instrument variables: (i) calibrated night-time light in the year 2001 at the district level, and (ii) standard deviation of night-time light across the settlements within a district. The first indicator is a proxy for economic activities within a district. However, the second indicator captures the variation within a district, i.e., whether there is a concentration or dispersion of economic activities at the sub-district level. Further, we have also used interaction terms among these indicators. These interaction terms capture inter-district variation with respect to night-time light, thereby highlighting that higher economic activity is associated with concentrated or dispersed development in a district.

The economic reasons for choosing the aforementioned exogenous instrument variables are as follows. First, the geographic distribution of colleges across the districts in India is based on the literature focusing on the supply-side effect of availability and accessibility to colleges leading to improved educational outcomes for the local residents. Card (1993) used the geographic variation in access to 4-year colleges as an instrument for years of education in the wage equation. Ideally, we would have liked to have the distance of households to the college in our model, but in the absence of such information, the availability of the number of colleges in the district is used for our analysis. This is in line with the study of Dee (2004) who used the availability of colleges in the county as an instrumental variable for accessibility to education while estimating the civic returns to education. Sekhri et al. (2022) applied the addition/construction of new colleges in the district as an instrument variable while estimating the impact of human capital for women. Other studies[23] along similar lines are Doyle and Skinner (2016), Kramer and Tamm (2018), Moretti (2014), and Rouse (1995).

Second, following the study of Henderson et al. (2011, 2012), a stream of literature has emerged, which uses the night-time light intensity to measure the differences in the geographic variation of the economic activities. In this literature, the night-time light is used as an instrument and proxy for urbanization (Henderson et al., 2011), poverty (Andreano et al., 2021; Asher et al., 2021), growth (Henderson et al., 2012), overall development (Castelló-Climent et al., 2018), and regional inequality (Singhal et al., 2020). In recent studies, night-time light is also used to explain the educational inequalities across the regions (Bruederle and Hodler, 2018; Qi et al., 2021). These studies have argued that the initial level of development and economic activities as measured by night-time light affects the access to education. It shows that regions with a lower level of human development have higher educational inequalities and vice-versa. We have applied the calibrated night-time light in 2001 as an indicator of average initial development in the district and the standard deviation of night-time light across the settlements in a district as a measure of the unequal distribution of this development. Bruederle and Hodler (2018), in the context of 29 African countries, observed that night-time light is positively associated with access to schools and years of schooling at the local level. From this perspective, night-time light has been used as instrumental variables in our Lewbel IV model.

---

[23] Please refer to Orosz et al. (2020) for a detailed discussion on a variety of instrumental variables being used in studies related to higher education.

## *Weak Instruments*

A major concern in IV models is to check whether the instruments are powerful enough to be used in the empirical exercise or are weak instruments. A weak instrument is weakly correlated to an endogenous regressor, which can make the estimation and inference unreliable. Andrews et al. (2019) using a survey of around 230 specifications in articles published in American Economic Review between 2014 and 2018 showed that the F-test statistic value of the first stage estimation in their Two-stage least squared (2SLS) IV models was very low. This led to their results being unreliable and sensitive to the sample. A solution to this concern is to use a hybrid instrument variable model that applies exogenous external instruments along with statistically constructed instruments to increase the explanatory power (F-test statistic) of the first stage estimation in the 2SLS IV model. Klein and Vella (2010) and Lewbel (2012) provided an approach to generate statistical instrument variables that can be combined with the exogenous instrument variables to resolve the problem of weak instruments. In this study, we have implemented this alternate approach. A number of studies have adopted this approach, for example, Millimet and Roy (2016) measured the effect of environmental regulations; Mishra and Smyth (2015) measured the returns to schooling; Dang and Rogers (2016) measured the effect of investment in children. For a detailed discussion on the implementation of this approach, refer to Baum and Lewbel (2019).

## *Empirical Model*

There are two Lewbel IV models that are estimated. The first model is:

$$log y_{ir} = \alpha + \beta_1 Edu_i + \beta_2 X_{1i} + \beta_2 X_{2r} + \beta_3 D_r + \beta_4 M_i + \lambda_i^{emp} + \lambda_i^{wage} + \lambda_i^{mig} + e_{ir}$$

$$Edu_{ir} = \gamma_o + \gamma_1 Z_{1r} + \gamma_{21} Z_{21r} + \gamma_{22} Z_{22r} + \gamma_3 Z_{3i} + \gamma_4 Z_{3i} * Z_{1r} + \gamma_5 G_{ir} + v_{ir} \ (5)$$

In the second equation, we have the first stage of modelling of the endogenous variable, i.e., attained years of education. Here $Z_{1r}$ is the number of colleges/educational institutions in the district; $Z_{21r}$ is log of calibrated total night-time light at the district level, and $Z_{21r}$ is the standard deviation of night-time light within the district; $Z_{3i}$ is the monthly per capita expenditure (MPCE) at the household level which is a proxy for household income status. Lastly, the variables, number of colleges and the household income status have been interacted in line with Card (1993). Additionally, to overcome the concerns of weak instruments, we have also included $G_i$, the statistically generated instrument variables in our model (Lewbel, 2012). In the first equation (1), the attained years of education variable has been replaced with the instrumental variables and analyzed using a two-stage least square approach (Wooldridge, 2010).

In the next model, three variables, i.e., required years of education, surplus years of education and deficit years of education have been replaced with the same set of instrumental variables as in the first model for analysis.

$$logy_{ir} = \alpha + \delta_1 Edu_i^r + \delta_2 Edu_i^s + \delta_3 Edu_i^d + \beta_2 X_{1i} + \beta_2 X_{2r} + \beta_3 D_r + + \beta_4 M_i + \lambda_i{}^{emp} + \lambda_i{}^{wage} + \lambda_i{}^{mig} + e_{ir}$$

$$Edu_i^r = \theta_o + \theta_1 Z_{1r} + \theta_{21} Z_{21r} + \theta_{22} Z_{22r} + \theta_3 Z_{3i} + \theta_4 Z_{3i} * Z_{1r} + \theta_5 G_{ir} + v1_{ir} \quad (6)$$

$$Edu_i^s = \theta_o + \theta_1 Z_{1r} + \theta_{21} Z_{21r} + \theta_{22} Z_{22r} + \theta_3 Z_{3i} + \theta_4 Z_{3i} * Z_{1r} + \theta_5 G_{ir} + v2_{ir} \quad (7)$$

$$Edu_i^d = \theta_o + \theta_1 Z_{1r} + \theta_{21} Z_{21r} + \theta_{22} Z_{22r} + \theta_3 Z_{3i} + \theta_4 Z_{3i} * Z_{1r} + \theta_5 G_{ir} + v3_{ir} \quad (8)$$

### *Diagnostic Tests for Lewbel IV Models*

The diagnostic tests of the Lewbel IV models for non-migrants and intra-district migrants are reported in Table A12. For the application of Lewbel IV models, the first requirement is the presence of heteroskedasticity while estimating the OLS model for the sample used in the analysis. The Breusch-Pagan/Cook-Weisberg test for heteroskedasticity suggests that there is heteroskedasticity in both the sub-samples, which makes it suitable for the Lewbel IV estimation. We have conducted the weak instrument test (Sanderson-Windmeijer (SW) test) for multiple instruments and rejected the null hypothesis that the instruments are jointly weak for estimation. Further, the under identification test (Kleibergen- Papp rk LM statistics) has also indicated that instruments used for the instrumental variable models are relevant. However, we have observed that Hansen J statistics is failing to reject the null hypothesis, thus suggesting that at least one instrument is redundant in our model.

**Section A4**

**Additional results for migrants with any reason for migration**

We have estimated the model for migrants by their reason for migration. The broad category of migrants by reason are: work related, for education, forced migration (due to natural disaster, socio-political problems and displacement for development projects), marriage, tied movers (due to mobility of parents/family members) and others. Before we go ahead with the results, due to the smaller number of observations at the national level for migration on account of education, forced migration and others, the findings for these categories should be taken with some caution.

As compared to the work-related migrants, migrants due to marriage and tied movers have lower returns to attained education (see Table A13a). This finding is not surprising because the migration is not optimally chosen but is the result of some non-labor market related decisions.

Further, we have observed that though forced and marriage migrants get higher returns to required education (see Table A13b), they also face higher penalty for deficit education and get relatively lower returns for surplus education. Thus, they experience double penalty in the labor market.

**Section A5**

**Selected results for only married migrants**

In this subsection, we discuss the findings for a selective sample consisting of only married migrants. The prime reason for this analysis is to check whether a spouse's education and employment status affect the returns to education and EOM at the destination labor market. The results are listed in Tables A14a and A14b. In this model, we have included all the other explanatory variables and two additional variables for examining a spouse's characteristics.

From Table A14a, we have found that married migrants have higher returns to attained education as compared to married non-migrant individuals. Coming to the returns to required education, there seems to be lower returns for the migrants than non-migrants, but migrants derive higher returns for surplus education. This indicates that migrants on account of spatial flexibility are able to improve their wages marginally. In terms of deficit education, there is no significant difference between migrants and non-migrants.

Coming to the effect of spouse's education, we have found that it leads to higher wages for the migrants as compared to non-migrants (see Table A14b). On the other hand, if your spouse is employed that puts a higher penalty for the migrants than on non-migrants. One explanation for this finding arises from the fact that migrants have to relocate keeping the employment of their spouse in mind due to which they get a penalty for their inflexibility in the labor market. They are supposed to find two jobs at the same place, which compels them to take a hit on their wages.

**Section A6**

**Return to Education and Education-Occupation Mismatch by Gender**

In this section, we analyse the returns to education for migrants by gender (see Table A15). We have found that females (4 percent) earn a higher return to education than males (3 percent). However, the results are required to be interpreted with caution as the proportion of females in work-related migrants is miniscule (11 percent) compared to males (89 percent). This finding is aligned to the literature that suggests while males are negatively selected to migrate, higher education increases migration among women (Kanaiaupuni, 2000). Besides, in the lower panel of Table A15, we have observed that while females earn higher returns to required education, returns to surplus and deficit years of education are statistically similar. Therefore, female migrants are better rewarded in the labor market as compared to male migrants. Also, the results for the overall group are inclined toward males owing to the higher proportion of males in the overall category.

| | Total | Migrants | Non-migrants |
|---|---|---|---|
| Table A0: Details about the sample | | | |
| Total Data | 572,254 | 169,445 | 402,809 |
| Sample Restrictions: | | | |
| Age-Individuals aged between 15-59 | 354,440 | 138,440 | 215,597 |
| Employment Status-Wage/Salary employed | 93,431 | 32,742 | 60,689 |
| Migration-Internal | | 32,438 | - |
| Migration-Work related | | 15,434 | - |
| Migration-Others reasons | | 16,897 | |
| Migration-No reason mentioned | | 107 | |

*Source:* Authors' calculation based on NSSO employment, unemployment and migration survey, 2007-08.

| | | Work-related migrants | Non-migrants |
|---|---|:---:|:---:|
| Table A1: Incidence of EOM and average daily wages (in Indian rupees) − Wage/Salary employed in the working-age group | | | |
| ***Overall*** | Proportion | 6 | 94 |
| ***Match status*** | Under | 15 | 12 |
| | Adequate | 70 | 71 |
| | Over | 15 | 18 |
| | **Total** | **100** | **100** |
| ***Wages (in Indian Rupees)*** | Under | 147 | 108 |
| | Adequate | 238 | 111 |
| | Over | 258 | 107 |

*Source:* Authors' calculation based on NSSO employment, unemployment and migration survey, 2007-08.
*Note:* Sampling weights have been used.

| | (1) | (2) | (3) |
|---|---|---|---|
| Table A2: Sample Selection Probit Model – in working-age group | | | |
| | Migration | Employment | Type of Employment |
| Years of Education | -0.00664*** | -0.0303*** | -0.00177** |
| | (0.000618) | (0.000712) | (0.000714) |
| Age | 0.0350*** | 0.223*** | 0.0177*** |
| | (0.00171) | (0.00185) | (0.00221) |
| **Gender (Base cat: Male)** | | | |
| Female | -0.0919*** | -0.991*** | 0.132*** |
| | (0.0117) | (0.00986) | (0.0185) |
| **Marital Status (Base cat: Unmarried)** | | | |
| Married | 0.128*** | 1.103*** | 0.0337*** |
| | (0.0109) | (0.0126) | (0.0118) |
| Others | 0.000822 | 0.185*** | 0.0251 |
| | (0.0322) | (0.0358) | (0.0300) |
| **Social Group (Base cat: Scheduled Tribe)** | | | |
| Scheduled Caste | 0.0951*** | -0.150*** | -0.143*** |
| | (0.0112) | (0.0123) | (0.0128) |
| OBC | 0.159*** | -0.174*** | 0.275*** |
| | (0.0104) | (0.0113) | (0.0115) |
| Others | 0.199*** | -0.287*** | 0.327*** |
| | (0.0111) | (0.0117) | (0.0125) |
| **Religion (Base cat: Hindu)** | | | |
| Muslim | -0.270*** | -0.0678*** | 0.186*** |
| | (0.00894) | (0.00903) | (0.0111) |
| Christian | -0.0368* | 0.0335* | 0.105*** |
| | (0.0212) | (0.0180) | (0.0194) |
| Others | -0.0362** | -0.0394** | 0.169*** |
| | (0.0168) | (0.0169) | (0.0199) |
| **Sector (Base cat: Rural)** | | | |
| Urban | 0.160*** | -0.381*** | 0.220*** |
| | (0.00703) | (0.00727) | (0.00862) |
| **Exclusion Variables** | | | |
| Migrant Network | 0.288*** | | |
| | (0.00644) | | |
| Migrant network Squared | -0.038*** | | |
| | (0.001) | | |
| Dependent members | | 0.0679*** | |
| | | (0.00271) | |
| **Household Type (Base cat: Self-employed)** | | | |
| regular wage/salary earning | | -0.0853*** | |
| | | (0.00959) | |
| casual labour | | 0.147*** | |

|  |  |
|---|---|
|  | (0.00739) |
| Others | -0.723*** |
|  | (0.0105) |
| Household Size | -0.0597*** |
|  | (0.00174) |

**Land possessed (area in hectare) (Base cat: Less than 0.005)**

|  |  |
|---|---|
| 0.005-0.01 | 0.126*** |
|  | (0.00997) |
| 0.02-0.20 | 0.227*** |
|  | (0.0112) |
| 0.21-0.40 | 0.609*** |
|  | (0.0126) |
| 0.41-1.00 | 1.200*** |
|  | (0.0122) |
| 1.01-2.00 | 1.563*** |
|  | (0.0142) |
| 2.01-3.00 | 1.755*** |
|  | (0.0225) |
| 3.01-4.00 | 1.808*** |
|  | (0.0338) |
| 4.01-6.00 | 1.905*** |
|  | (0.0423) |
| 6.01-8.00 | 2.179*** |
|  | (0.0732) |
| greater than 8.00 | 1.960*** |
|  | (0.0555) |

Source: Authors' calculation based on NSSO employment, unemployment and migration survey, 2007-08.
Note: (i) *** signals significant at 1% level, ** signals significant at 5% level, and * signals significant at 10% level.
(ii) Robust standard errors are in parenthesis.
(iii) The analysis also controls age squared, interaction of gender and marital status, and 35 states and union territories.

| | | Non-Migrants | Work-Related Migrants |
|---|---|---|---|
| | Table A3: Summary statistics for migrants and non-migrants in working-age group | | |
| | Years of Education | 7.69 | 9.40 |
| | Required Years of Education | 6.14 | 8.71 |
| | Surplus Years of Education | 1.99 | 1.67 |
| | Deficit Years of Education | 1.67 | 1.30 |
| | Age | 31.14 | 36.44 |
| *Gender* | Male | 72.01 | 90.66 |
| | Female | 27.99 | 9.34 |
| *Marital Status* | Unmarried | 38.01 | 17.71 |
| | Married | 58.98 | 79.67 |
| | Others | 3.01 | 2.62 |
| *Social Group* | Scheduled Tribe | 9.38 | 4.81 |
| | Scheduled Caste | 19.01 | 15.2 |
| | OBC | 41.88 | 34.51 |
| | Others | 29.73 | 45.47 |
| *Religion* | Hinduism | 81.05 | 85.79 |
| | Islamism | 13.57 | 8.71 |
| | Christianity | 2.56 | 2.8 |
| | Others | 2.83 | 2.7 |
| *Sector* | Rural | 73.66 | 24.06 |
| | Urban | 26.34 | 75.94 |

*Source:* Authors' calculation based on NSSO employment, unemployment and migration survey, 2007-08.
*Note*: Sampling weights have been used.

Table A4a: Incidence and average daily wages (in Indian rupees) of work-related migrants in the working age-group – by reason to migrate

| | Overall proportion | Proportion | | | Wages (in Indian Rupees) | | |
|---|---|---|---|---|---|---|---|
| | | Under | Adequate | Over | Under | Adequate | Over |
| Job search | 68 | 18 | 69 | 13 | 125 | 165 | 182 |
| Take-up job | 32 | 14 | 70 | 16 | 149 | 269 | 317 |

*Source:* Authors' calculation based on NSSO employment, unemployment and migration survey, 2007-08.
*Note:* Sampling weights have been used.

Table A4b: Incidence and average daily wages (in Indian rupees) of work-related migrants in the working age-group – by gender

| | Overall proportion | Proportion | | | Wages (in Indian Rupees) | | |
|---|---|---|---|---|---|---|---|
| | | Under | Adequate | Over | Under | Adequate | Over |
| Male | 91 | 14 | 71 | 15 | 155 | 240 | 260 |
| Female | 9 | 21 | 68 | 11 | 95 | 208 | 244 |

*Source:* Authors' calculation based on NSSO employment, unemployment and migration survey, 2007-08.
*Note:* Sampling weights have been used.

Table A4c: Incidence and average daily wages (in Indian rupees) of work-related migrants in the working age-group – by distance

| | Overall proportion | Proportion | | | Wages (in Indian Rupees) | | |
|---|---|---|---|---|---|---|---|
| | | Under | Adequate | Over | Under | Adequate | Over |
| Rural-Rural | 18 | 15 | 72 | 13 | 137 | 154 | 160 |
| Rural-Urban | 50 | 18 | 69 | 13 | 138 | 196 | 203 |
| Urban-Rural | 6 | 13 | 71 | 17 | 177 | 216 | 195 |
| Urban-Urban | 25 | 9 | 72 | 19 | 189 | 381 | 392 |
| Intra-district | 26 | 14 | 72 | 14 | 149 | 208 | 216 |
| Inter-district | 37 | 15 | 69 | 16 | 150 | 271 | 249 |

*Source:* Authors' calculation based on NSSO employment, unemployment and migration survey, 2007-08.
*Note:* Sampling weights have been used.

Table A4d: Incidence and average daily wages (in Indian rupees) of work-related migrants in the working age-group – by years since migration

| | Overall proportion | Proportion | | | Wages (in Indian Rupees) | | |
|---|---|---|---|---|---|---|---|
| | | Under | Adequate | Over | Under | Adequate | Over |
| 0-2 | 24 | 14 | 69 | 16 | 133 | 220 | 185 |
| 3-5 | 20 | 13 | 70 | 17 | 135 | 226 | 204 |
| 6-10 | 21 | 15 | 72 | 13 | 138 | 219 | 238 |
| 11 or above | 35 | 17 | 70 | 12 | 174 | 277 | 418 |

*Source:* Authors' calculation based on NSSO employment, unemployment and migration survey, 2007-08.
*Note:* Sampling weights have been used.

Table A5a: Returns to education for wage/salary employed in working-age group: Work-related migrants and non-migrants

| | Dependent variable: ln(wage) | |
|---|---|---|
| | (1) | (2) |
| Explanatory variables | Work related migrants | Non-migrants |
| Attained education | 0.0298*** | 0.0244*** |
| | (0.00182) | (0.000937) |
| Age | 0.0108** | -0.00788*** |
| | (0.00519) | (0.00298) |
| Age squared | 3.16e-06 | 0.000227*** |
| | (6.50e-05) | (3.84e-05) |
| **Gender (base cat: male)** | -0.206*** | -0.214*** |
| Female | (0.0507) | (0.0241) |
| | | |
| **Marital status (base cat: unmarried male)** | | |
| Male*married | 0.0948*** | 0.0465*** |
| | (0.0248) | (0.0143) |
| Male*Others | 0.0448 | -0.0421* |
| | (0.0544) | (0.0248) |
| **Gender*marital status** | | |
| Female*married | -0.203*** | -0.335*** |
| | (0.0710) | (0.0428) |
| Female*others | -0.160* | -0.243*** |
| | (0.0868) | (0.0466) |
| **Social group (base cat: Scheduled tribe)** | | |
| Scheduled caste | -0.0343 | -0.00376 |
| | (0.0297) | (0.0152) |
| Other Backward Class (OBC) | -0.0165 | -0.0149 |
| | (0.0287) | (0.0158) |
| Others | 0.0264 | 0.0471** |
| | (0.0297) | (0.0186) |
| **Religion (base cat: Hindu)** | | |
| Muslims | -0.0512** | 0.0445*** |
| | (0.0256) | (0.0147) |
| Christians | 0.0171 | 0.0755*** |
| | (0.0311) | (0.0271) |
| Others | 0.0541 | 0.0550** |
| | (0.0478) | (0.0268) |
| **Other controls** | | |
| Sector (rural/urban) | Y | Y |
| Origin location (state) | Y | N |
| Destination location (state) | Y | Y |
| Industry category (1 digit) | Y | Y |
| Occupation category (1 digit) | Y | Y |
| Reason for migration | Y | N |
| Migration stream | Y | N |
| Migration distance | Y | N |
| Years since migration | Y | N |
| International immigration rate (state level) | Y | Y |
| ln(working age population) | Y | Y |
| ln(labour force) | Y | Y |
| Employment status (before migration) | Y | N |
| **Sample selection corrections** | | |
| Inverse mills ratio- employment | Y | Y |
| Inverse mills ratio- occupation | Y | Y |
| Inverse mills ratio- migration | Y | N |
| | | |
| Clustered standard errors (at first stage unit) | Y | Y |
| Observations | 13,887 | 56,440 |
| R-squared | 0.501 | 0.359 |

Source: Authors' calculation based on NSSO employment, unemployment and migration survey, 2007-08.
Note:
(i) *** signals significant at 1% level, ** signals significant at 5% level and * signals significant at 10% level.
(ii) robust standard errors are given in parenthesis.

Table A5b: Returns to education for wage/salary employed in working-age group: Work-related migrants and non-migrants

| Explanatory variables | Dependent variable: ln(wage) | |
|---|---|---|
| | (1) | (2) |
| | Work related migrants | Non-migrants |
| Required education | 0.0751*** | 0.0764*** |
| | (0.00584) | (0.00477) |
| Surplus education | 0.0324*** | 0.0227*** |
| | (0.00334) | (0.00194) |
| Deficit education | -0.0241*** | -0.0238*** |
| | (0.00357) | (0.00194) |
| Age | 0.0106** | -0.00646** |
| | (0.00514) | (0.00300) |
| Age squared | 5.82e-07 | 0.000209*** |
| | (6.46e-05) | (3.86e-05) |
| **Gender (base cat: male)** | | |
| Female | -0.192*** | -0.218*** |
| | (0.0504) | (0.0245) |
| **Marital status (base cat: unmarried male)** | | |
| Male*married | 0.0968*** | 0.0524*** |
| | (0.0246) | (0.0145) |
| Male*Others | 0.0433 | -0.0422* |
| | (0.0537) | (0.0256) |
| **Gender*marital status** | | |
| Female*married | -0.167** | -0.301*** |
| | (0.0703) | (0.0435) |
| Female*others | -0.125 | -0.209*** |
| | (0.0859) | (0.0472) |
| **Social group (base cat: Scheduled tribe)** | | |
| Scheduled caste | -0.0272 | 0.00458 |
| | (0.0295) | (0.0155) |
| Other Backward Class (OBC) | -0.0128 | -0.00942 |
| | (0.0289) | (0.0161) |
| Others | 0.0300 | 0.0516*** |
| | (0.0298) | (0.0190) |
| **Religion (base cat: Hindu)** | | |
| Muslims | -0.0610** | 0.0431*** |
| | (0.0256) | (0.0149) |
| Christians | 0.0287 | 0.0817*** |
| | (0.0300) | (0.0279) |
| Others | 0.0640 | 0.0406 |
| | (0.0465) | (0.0271) |
| **Other controls** | | |
| Sector (rural/urban) | Y | Y |
| Origin location (state) | Y | N |
| Destination location (state) | Y | Y |
| Industry category (1 digit) | Y | Y |
| Occupation category (1 digit) | Y | Y |
| Reason for migration | Y | N |
| Migration stream | Y | N |
| Migration distance | Y | N |
| Years since migration | Y | N |
| International immigration rate (state level) | Y | Y |
| ln(working age population) | Y | Y |
| ln(labour force) | Y | Y |
| Employment status (before migration) | Y | N |
| **Sample selection corrections** | | |
| Inverse mills ratio- employment | Y | Y |
| Inverse mills ratio- occupation | Y | Y |
| Inverse mills ratio- migration | Y | N |
| | | |
| Clustered standard errors (at first stage unit) | Y | Y |
| Observations | 13,745 | 53,824 |
| R-squared | 0.512 | 0.375 |

Source: Authors' calculation based on NSSO employment, unemployment and migration survey, 2007-08.
Note:
(i) *** signals significant at 1% level, ** signals significant at 5% level and * signals significant at 10% level. (ii) robust standard errors are given in parenthesis.

Table A6a: Returns to education and EOM for employed migrants in working-age group: by reason to migrate (work related)

| | Dependent variable: ln(wage) | | | |
|---|---|---|---|---|
| | (1) | (2) | (3) | (4) |
| Explanatory variables | Overall | Confirm | Job search | Others |
| Attained education | 0.0298*** | 0.0204*** | 0.0337*** | 0.0433*** |
| | (0.00182) | (0.00230) | (0.00325) | (0.00561) |
| Age | 0.0108** | 0.00954 | 0.0102 | 0.0272** |
| | (0.00519) | (0.00755) | (0.0101) | (0.0129) |
| Age squared | 3.16e-06 | -4.05e-05 | 5.84e-07 | -0.000134 |
| | (6.50e-05) | (9.69e-05) | (0.000127) | (0.000154) |
| **Gender (base cat: male)** | | | | |
| Female | -0.206*** | -0.101 | -0.174** | -0.176 |
| | (0.0507) | (0.0920) | (0.0711) | (0.107) |
| **Marital status (base cat: unmarried male)** | | | | |
| Male*married | 0.0948*** | 0.0248 | 0.150*** | 0.129 |
| | (0.0248) | (0.0342) | (0.0409) | (0.0803) |
| Male*Others | 0.0448 | 0.0429 | 0.0606 | 0.102 |
| | (0.0544) | (0.0866) | (0.106) | (0.0916) |
| **Gender*marital status** | | | | |
| Female*married | -0.203*** | -0.174 | -0.238** | -0.397** |
| | (0.0710) | (0.110) | (0.120) | (0.187) |
| Female*others | -0.160* | -0.224 | -0.176 | -0.316* |
| | (0.0868) | (0.137) | (0.154) | (0.167) |
| **Social group (base cat: Scheduled tribe)** | | | | |
| Scheduled caste | -0.0343 | 0.0374 | -0.0259 | -0.138** |
| | (0.0297) | (0.0415) | (0.0591) | (0.0562) |
| Other Backward Class (OBC) | -0.0165 | 0.0390 | 0.00241 | -0.0972* |
| | (0.0287) | (0.0397) | (0.0570) | (0.0565) |
| Others | 0.0264 | 0.120*** | 0.0471 | -0.128** |
| | (0.0297) | (0.0421) | (0.0585) | (0.0573) |
| **Religion (base cat: Hindu)** | | | | |
| Muslims | -0.0512** | -0.0733** | -0.0640 | -0.0429 |
| | (0.0256) | (0.0356) | (0.0458) | (0.0473) |
| Christians | 0.0171 | 0.0803 | -0.0792 | -0.00725 |
| | (0.0311) | (0.0562) | (0.0522) | (0.0505) |
| Others | 0.0541 | 0.0215 | 0.0630 | 0.0861 |
| | (0.0478) | (0.0788) | (0.0851) | (0.0636) |
| **Other controls** | | | | |
| Sector (rural/urban) | Y | Y | Y | Y |
| Origin location (state) | Y | Y | Y | Y |
| Destination location (state) | Y | Y | Y | Y |
| Industry category (1 digit) | Y | Y | Y | Y |
| Occupation category (1 digit) | Y | Y | Y | Y |
| Reason for migration | Y | N | N | N |
| Migration stream | Y | Y | Y | Y |
| Migration distance | Y | Y | Y | Y |
| Years since migration | Y | Y | Y | Y |
| International immigration rate (state level) | Y | Y | Y | Y |
| ln(working age population) | Y | Y | Y | Y |
| ln(labour force) | Y | Y | Y | Y |
| Employment status (before migration) | Y | Y | Y | Y |
| **Sample selection corrections** | | | | |
| Inverse mills ratio- employment | Y | Y | Y | Y |
| Inverse mills ratio- occupation | Y | Y | Y | Y |
| Inverse mills ratio- migration | Y | Y | Y | Y |
| | | | | |
| Clustered standard errors (at first stage unit) | Y | Y | Y | Y |
| Observations | 13,887 | 6,635 | 4,104 | 3,148 |
| R-squared | 0.501 | 0.369 | 0.482 | 0.377 |

Source: Authors' calculation based on NSSO employment, unemployment and migration survey, 2007-08.
Note:
(i) *** signals significant at 1% level, ** signals significant at 5% level and * signals significant at 10% level.
(ii) robust standard errors are given in parenthesis.

Table A6b: Returns to education and EOM for employed migrants in working-age group: by reason to migrate (work related)

| Explanatory variables | Dependent variable: ln(wage) | | | |
|---|---|---|---|---|
| | (1) Overall | (2) Confirm | (3) Job search | (4) Others |
| Required education | 0.0751*** | 0.0648*** | 0.0787*** | 0.0770*** |
| | (0.00584) | (0.0104) | (0.00918) | (0.00913) |
| Surplus education | 0.0324*** | 0.0255*** | 0.0304*** | 0.0375*** |
| | (0.00334) | (0.00462) | (0.00619) | (0.00744) |
| Deficit education | -0.0241*** | -0.0153*** | -0.0314*** | -0.0463*** |
| | (0.00357) | (0.00389) | (0.00604) | (0.0163) |
| Age | 0.0106** | 0.00857 | 0.0114 | 0.0263** |
| | (0.00514) | (0.00749) | (0.00999) | (0.0129) |
| Age squared | 5.82e-07 | -3.15e-05 | -1.82e-05 | -0.000130 |
| | (6.46e-05) | (9.68e-05) | (0.000126) | (0.000154) |
| **Gender (base cat: male)** | | | | |
| Female | -0.192*** | -0.0881 | -0.161** | -0.151 |
| | (0.0504) | (0.0913) | (0.0704) | (0.107) |
| **Marital status (base cat: unmarried male)** | | | | |
| Male*married | 0.0968*** | 0.0214 | 0.157*** | 0.128 |
| | (0.0246) | (0.0337) | (0.0408) | (0.0803) |
| Male*Others | 0.0433 | 0.0260 | 0.0672 | 0.114 |
| | (0.0537) | (0.0858) | (0.103) | (0.0941) |
| **Gender*marital status** | | | | |
| Female*married | -0.167** | -0.151 | -0.199* | -0.349* |
| | (0.0703) | (0.108) | (0.118) | (0.187) |
| Female*others | -0.125 | -0.186 | -0.146 | -0.283* |
| | (0.0859) | (0.135) | (0.152) | (0.167) |
| **Social group (base cat: Scheduled tribe)** | | | | |
| Scheduled caste | -0.0272 | 0.0485 | -0.0209 | -0.138** |
| | (0.0295) | (0.0411) | (0.0593) | (0.0561) |
| Other Backward Class (OBC) | -0.0128 | 0.0447 | 0.0104 | -0.101* |
| | (0.0289) | (0.0404) | (0.0570) | (0.0565) |
| Others | 0.0300 | 0.122*** | 0.0519 | -0.119** |
| | (0.0298) | (0.0426) | (0.0586) | (0.0575) |
| **Religion (base cat: Hindu)** | | | | |
| Muslims | -0.0610** | -0.0792** | -0.0775* | -0.0514 |
| | (0.0256) | (0.0354) | (0.0461) | (0.0470) |
| Christians | 0.0287 | 0.0952* | -0.0543 | -0.00190 |
| | (0.0300) | (0.0540) | (0.0509) | (0.0504) |
| Others | 0.0640 | 0.0463 | 0.0741 | 0.0873 |
| | (0.0465) | (0.0760) | (0.0851) | (0.0640) |
| **Other controls** | | | | |
| Sector (rural/urban) | Y | Y | Y | Y |
| Origin location (state) | Y | Y | Y | Y |
| Destination location (state) | Y | Y | Y | Y |
| Industry category (1 digit) | Y | Y | Y | Y |
| Occupation category (1 digit) | Y | Y | Y | Y |
| Reason for migration | N | N | N | N |
| Migration stream | Y | Y | Y | Y |
| Migration distance | Y | Y | Y | Y |
| Years since migration | Y | Y | Y | Y |
| International immigration rate (state level) | Y | Y | Y | Y |
| ln(working age population) | Y | Y | Y | Y |
| ln(labour force) | Y | Y | Y | Y |
| Employment status (before migration) | Y | Y | Y | Y |
| **Sample selection corrections** | | | | |
| Inverse mills ratio- employment | Y | Y | Y | Y |
| Inverse mills ratio- occupation | Y | Y | Y | Y |
| Inverse mills ratio- migration | Y | Y | Y | Y |
| | | | | |
| Clustered standard errors (at first stage unit) | Y | Y | Y | Y |
| Observations | 13,745 | 6,541 | 4,071 | 3,133 |
| R-squared | 0.512 | 0.384 | 0.490 | 0.388 |

Source: Authors' calculation based on NSSO employment, unemployment and migration survey, 2007-08.
Note: (i) *** signals significant at 1% level, ** signals significant at 5% level and * signals significant at 10% level.
(ii) robust standard errors are given in parenthesis.

Table A7a: Returns to education and EOM for wage/salary employed migrants in working-age group: by migration stream

| Explanatory variables | Dependent variable: ln(wage) | | | | |
|---|---|---|---|---|---|
| | (1) Overall | (2) Rural-rural | (3) Rural-urban | (4) Urban-rural | (5) Urban-urban |
| Attained education | 0.0298*** | 0.0242*** | 0.0241*** | 0.0305*** | 0.0460*** |
| | (0.00182) | (0.00307) | (0.00261) | (0.00639) | (0.00478) |
| Age | 0.0108** | -0.00455 | 0.0250*** | -0.0132 | 0.0201 |
| | (0.00519) | (0.0101) | (0.00836) | (0.0158) | (0.0156) |
| Age squared | 3.16e-06 | 0.000176 | -0.000183* | 0.000297 | -7.89e-05 |
| | (6.50e-05) | (0.000129) | (0.000105) | (0.000199) | (0.000185) |
| **Gender (base cat: male)** | | | | | |
| Female | -0.206*** | -0.227** | -0.323*** | -0.110 | -0.150 |
| | (0.0507) | (0.103) | (0.0873) | (0.163) | (0.138) |
| **Marital status (base cat: unmarried male)** | | | | | |
| Male*married | 0.0948*** | 0.0922* | 0.142*** | 0.140* | 0.0185 |
| | (0.0248) | (0.0494) | (0.0404) | (0.0767) | (0.0731) |
| Male*Others | 0.0448 | 0.0211 | 0.0883 | 0.249 | 0.0517 |
| | (0.0544) | (0.0910) | (0.0948) | (0.177) | (0.122) |
| Gender*marital status | | | | | |
| Female*married | -0.203*** | -0.371** | -0.280** | -0.289 | -0.213 |
| | (0.0710) | (0.147) | (0.111) | (0.203) | (0.177) |
| Female*others | -0.160* | -0.320** | -0.129 | -0.339 | -0.238 |
| | (0.0868) | (0.159) | (0.142) | (0.282) | (0.179) |
| **Social group (base cat: Scheduled tribe)** | | | | | |
| Scheduled caste | -0.0343 | 0.00221 | -0.0733 | -0.132* | -0.0837 |
| | (0.0297) | (0.0460) | (0.0494) | (0.0777) | (0.0618) |
| Other Backward Class (OBC) | -0.0165 | -0.0261 | -0.0607 | -0.0571 | -0.130** |
| | (0.0287) | (0.0475) | (0.0478) | (0.0734) | (0.0561) |
| Others | 0.0264 | 0.0197 | -0.0418 | -0.0728 | -0.0852 |
|  | (0.0297) | (0.0508) | (0.0493) | (0.0845) | (0.0585) |
| **Religion (base cat: Hindu)** |  |  |  |  |  |
| Muslims | -0.0512** | 0.0154 | -0.0281 | 0.0309 | -0.0682 |
|  | (0.0256) | (0.0533) | (0.0367) | (0.103) | (0.0531) |
| Christians | 0.0171 | -0.00384 | 0.00218 | -0.0202 | -0.0122 |
|  | (0.0311) | (0.0598) | (0.0468) | (0.0931) | (0.0555) |
| Others | 0.0541 | 0.0294 | 0.0326 | 0.267* | -0.0333 |
|  | (0.0478) | (0.0754) | (0.0762) | (0.148) | (0.0889) |
| **Other controls** |  |  |  |  |  |
| Sector (rural/urban) | Y | N | N | N | N |
| Origin location (state) | Y | Y | Y | Y | Y |
| Destination location (state) | Y | Y | Y | Y | Y |
| Industry category (1 digit) | Y | Y | Y | Y | Y |
| Occupation category (1 digit) | Y | Y | Y | Y | Y |
| Reason for migration | Y | Y | Y | Y | Y |
| Migration stream | Y | N | N | N | N |
| Migration distance | Y | Y | Y | Y | Y |
| Years since migration | Y | Y | Y | Y | Y |
| International immigration rate (state level) | Y | Y | Y | Y | Y |
| ln(working age population) | Y | Y | Y | Y | Y |
| ln(labour force) | Y | Y | Y | Y | Y |
| Employment status (before migration) | Y | Y | Y | Y | Y |
| **Sample selection corrections** |  |  |  |  |  |
| Inverse mills ratio- employment | Y | Y | Y | Y | Y |
| Inverse mills ratio- occupation | Y | Y | Y | Y | Y |
| Inverse mills ratio- migration | Y | Y | Y | Y | Y |
|  |  |  |  |  |  |
| Clustered standard errors (at first stage unit) | Y | Y | Y | Y | Y |
| Observations | 13,887 | 3,937 | 5,715 | 1,086 | 3,149 |
| R-squared | 0.501 | 0.498 | 0.424 | 0.587 | 0.505 |

Table A7b: Returns to education and EOM for wage/salary employed migrants in working-age group: by migration stream

Dependent variable: ln(wage)

| Explanatory variables | (1) Overall | (2) Rural-rural | (3) Rural-urban | (4) Urban-rural | (5) Urban-urban |
|---|---|---|---|---|---|
| Required education | 0.0751*** | 0.0699*** | 0.0725*** | 0.0936*** | 0.0750*** |
| | (0.00584) | (0.0115) | (0.00915) | (0.0170) | (0.0110) |
| Surplus education | 0.0324*** | 0.0222*** | 0.0332*** | 0.0233** | 0.0486*** |
| | (0.00334) | (0.00645) | (0.00505) | (0.0108) | (0.00762) |
| Deficit education | -0.0241*** | -0.0230*** | -0.0140*** | -0.0322** | -0.0402*** |
| | (0.00357) | (0.00647) | (0.00384) | (0.0162) | (0.0121) |
| Age | 0.0106** | -0.00649 | 0.0251*** | -0.0117 | 0.0184 |
| | (0.00514) | (0.00973) | (0.00826) | (0.0156) | (0.0156) |
| Age squared | 5.82e-07 | 0.000197 | -0.000192* | 0.000274 | -6.64e-05 |
| | (6.46e-05) | (0.000126) | (0.000103) | (0.000196) | (0.000187) |
| **Gender (base cat: male)** | | | | | |
| Female | -0.192*** | -0.231** | -0.307*** | -0.0710 | -0.132 |
| | (0.0504) | (0.1000) | (0.0855) | (0.165) | (0.137) |
| **Marital status (base cat: unmarried male)** | | | | | |
| Male*married | 0.0968*** | 0.104** | 0.139*** | 0.142* | 0.0214 |
| | (0.0246) | (0.0483) | (0.0401) | (0.0786) | (0.0730) |
| Male*Others | 0.0433 | 0.0396 | 0.0521 | 0.252 | 0.0814 |
| | (0.0537) | (0.0906) | (0.0905) | (0.165) | (0.123) |
| **Gender*marital status** | | | | | |
| Female*married | -0.167** | -0.310** | -0.274** | -0.189 | -0.134 |
| | (0.0703) | (0.149) | (0.109) | (0.206) | (0.171) |
| Female*others | -0.125 | -0.272* | -0.0893 | -0.248 | -0.212 |
| | (0.0859) | (0.159) | (0.137) | (0.279) | (0.179) |
| **Social group (base cat: Scheduled tribe)** | | | | | |
| Scheduled caste | -0.0272 | 0.00511 | -0.0592 | -0.115 | -0.0763 |
| | (0.0295) | (0.0467) | (0.0479) | (0.0784) | (0.0617) |
| Other Backward Class (OBC) | -0.0128 | -0.0253 | -0.0624 | -0.0434 | -0.112** |
| | (0.0289) | (0.0482) | (0.0473) | (0.0720) | (0.0554) |
| Others | 0.0300 | 0.0132 | -0.0411 | -0.0421 | -0.0640 |
| | (0.0298) | (0.0519) | (0.0485) | (0.0850) | (0.0574) |
| **Religion (base cat: Hindu)** | | | | | |
| Muslims | -0.0610** | 0.00982 | -0.0404 | -0.0198 | -0.0794 |
| | (0.0256) | (0.0541) | (0.0364) | (0.104) | (0.0528) |
| Christians | 0.0287 | 0.0212 | -0.00584 | 0.0325 | -0.0101 |
| | (0.0300) | (0.0589) | (0.0466) | (0.0864) | (0.0548) |
| Others | 0.0640 | 0.0161 | 0.0938 | 0.254* | -0.0385 |
| | (0.0465) | (0.0757) | (0.0651) | (0.145) | (0.0879) |
| **Other controls** | | | | | |
| Sector (rural/urban) | Y | N | N | N | N |
| Origin location (state) | Y | Y | Y | Y | Y |
| Destination location (state) | Y | Y | Y | Y | Y |
| Industry category (1 digit) | Y | Y | Y | Y | Y |
| Occupation category (1 digit) | Y | Y | Y | Y | Y |
| Reason for migration | Y | Y | Y | Y | Y |
| Migration stream | Y | N | N | N | N |
| Migration distance | Y | Y | Y | Y | Y |
| Years since migration | Y | Y | Y | Y | Y |
| International immigration rate (state level) | Y | Y | Y | Y | Y |
| ln(working age population) | Y | Y | Y | Y | Y |
| ln(labour force) | Y | Y | Y | Y | Y |
| Employment status (before migration) | Y | Y | Y | Y | Y |
| **Sample selection corrections** | | | | | |
| Inverse mills ratio- employment | Y | Y | Y | Y | Y |
| Inverse mills ratio- occupation | Y | Y | Y | Y | Y |
| Inverse mills ratio- migration | Y | Y | Y | Y | Y |
| Clustered standard errors (at first stage unit) | Y | Y | Y | Y | Y |
| Observations | 13,745 | 3,865 | 5,680 | 1,064 | 3,136 |
| R-squared | 0.512 | 0.501 | 0.441 | 0.601 | 0.517 |

Table A8a: Returns to education and EOM for wage/salary employed migrants in working-age group: by distance

| Explanatory variables | Dependent variable: ln(wage) | | | |
|---|---|---|---|---|
| | (1) Overall | (2) Intra-district | (3) Inter-district (within state) | (4) Inter-state |
| Attained education | 0.0298*** | 0.028*** | 0.0340*** | 0.0270*** |
| | (0.00182) | (0.0033) | (0.00324) | (0.00268) |
| Age | 0.0108** | 0.00493 | 0.0194** | -7.51e-05 |
| | (0.00519) | (0.00922) | (0.00947) | (0.00883) |
| Age squared | 3.16e-06 | 7.07e-05 | -7.51e-05 | 0.000104 |
| | (6.50e-05) | (0.00012) | (0.00012) | (0.0001) |
| **Gender (base cat: male)** | | | | |
| Female | -0.206*** | -0.220** | -0.103 | -0.208** |
| | (0.0507) | (0.107) | (0.0824) | (0.0844) |
| **Marital status (base cat: unmarried male)** | | | | |
| Male*married | 0.0948*** | 0.220*** | 0.0905** | -0.0268 |
| | (0.0248) | (0.0579) | (0.0397) | (0.0360) |
| Male*Others | 0.0448 | 0.201** | 0.0110 | -0.0700 |
| | (0.0544) | (0.0813) | (0.113) | (0.104) |
| **Gender*marital status** | | | | |
| Female*married | -0.203*** | -0.527*** | -0.202* | 0.101 |
| | (0.0710) | (0.138) | (0.113) | (0.118) |
| Female*others | -0.160* | -0.343** | -0.257 | 0.186 |
| | (0.0868) | (0.146) | (0.159) | (0.163) |
| **Social group (base cat: Scheduled tribe)** | | | | |
| Scheduled caste | -0.0343 | -0.0111 | -0.0673 | 0.0253 |
| | (0.0297) | (0.0443) | (0.0496) | (0.0635) |
| Other Backward Class (OBC) | -0.0165 | 0.00659 | -0.0740 | 0.0385 |
| | (0.0287) | (0.0463) | (0.0455) | (0.0647) |
| Others | 0.0264 | 0.0402 | -0.0295 | 0.0865 |
| | (0.0297) | (0.0501) | (0.0461) | (0.0655) |
| **Religion (base cat: Hindu)** | | | | |
| Muslims | -0.0512** | 0.0272 | -0.105** | -0.0482 |
| | (0.0256) | (0.0485) | (0.0469) | (0.0419) |
| Christians | 0.0171 | 0.0286 | -0.0629 | -0.0113 |
| | (0.0311) | (0.0516) | (0.0403) | (0.0839) |
| Others | 0.0541 | 0.0919 | -0.0352 | 0.131* |
| | (0.0478) | (0.0893) | (0.0683) | (0.0700) |
| **Other controls** | | | | |
| Sector (rural/urban) | Y | Y | Y | Y |
| Origin location (state) | Y | N | N | Y |
| Destination location (state) | Y | Y | Y | Y |
| Industry category (1 digit) | Y | Y | Y | Y |
| Occupation category (1 digit) | Y | Y | Y | Y |
| Reason for migration | Y | Y | Y | Y |
| Migration stream | Y | Y | Y | Y |
| Migration distance | Y | N | N | N |
| Years since migration | Y | Y | Y | Y |
| International immigration rate (state level) | Y | Y | Y | Y |
| ln(working age population) | Y | Y | Y | Y |
| ln(labour force) | Y | Y | Y | Y |
| Employment status (before migration) | Y | Y | Y | Y |
| **Sample selection corrections** | | | | |
| Inverse mills ratio- employment | Y | Y | Y | Y |
| Inverse mills ratio- occupation | Y | Y | Y | Y |
| Inverse mills ratio- migration | Y | Y | Y | Y |
| | | | | |
| Clustered standard errors (at first stage unit) | Y | Y | Y | Y |
| Observations | 13,887 | 4,592 | 4,986 | 4,309 |
| R-squared | 0.501 | 0.530 | 0.530 | 0.451 |

Source: Authors' calculation based on NSSO employment, unemployment and migration survey, 2007-08.
Note:
(i) *** signals significant at 1% level, ** signals significant at 5% level and * signals significant at 10% level.
(ii) robust standard errors are given in parenthesis.

Table A8b: Returns to education and EOM for wage/salary employed migrants in working-age group: by distance

| Explanatory variables | Dependent variable: ln(wage) | | | |
|---|---|---|---|---|
| | (1) | (2) | (3) | (4) |
| | Overall | Intra-district | Inter-district (within state) | Inter-state |
| Required education | 0.0751*** | 0.0874*** | 0.0821*** | 0.0531*** |
| | (0.0059) | (0.01) | (0.0094) | (0.009) |
| Surplus education | 0.0324*** | 0.0361*** | 0.0291*** | 0.0325*** |
| | (0.003) | (0.0055) | (0.0059) | (0.00553) |
| Deficit education | -0.0241*** | -0.0166** | -0.0338*** | -0.0212*** |
| | (0.00357) | (0.00690) | (0.00621) | (0.00489) |
| Age | 0.0106** | 0.00843 | 0.0190** | -0.00392 |
| | (0.00514) | (0.00912) | (0.00946) | (0.00845) |
| Age squared | 5.82e-07 | 2.40e-05 | -8.01e-05 | 0.000145 |
| | (6.46e-05) | (0.000114) | (0.000116) | (0.000111) |
| **Gender (base cat: male)** | | | | |
| Female | -0.192*** | -0.211** | -0.0880 | -0.191** |
| | (0.0504) | (0.106) | (0.0814) | (0.0844) |
| **Marital status (base cat: unmarried male)** | | | | |
| Male*married | 0.0968*** | 0.219*** | 0.101** | -0.0237 |
| | (0.0246) | (0.0577) | (0.0395) | (0.0357) |
| Male*Others | 0.0433 | 0.183** | 0.0290 | -0.0736 |
| | (0.0537) | (0.0807) | (0.111) | (0.0999) |
| **Gender*marital status** | | | | |
| Female*married | -0.167** | -0.477*** | -0.172 | 0.131 |
| | (0.0703) | (0.135) | (0.112) | (0.118) |
| Female*others | -0.125 | -0.281* | -0.231 | 0.216 |
| | (0.0859) | (0.145) | (0.156) | (0.158) |
| **Social group (base cat: Scheduled tribe)** | | | | |
| Scheduled caste | -0.0272 | -0.00149 | -0.0602 | 0.0241 |
| | (0.0295) | (0.0440) | (0.0491) | (0.0634) |
| Other Backward Class (OBC) | -0.0128 | 0.00555 | -0.0586 | 0.0309 |
| | (0.0289) | (0.0469) | (0.0450) | (0.0647) |
| Others | 0.0300 | 0.0381 | -0.0117 | 0.0787 |
| | (0.0298) | (0.0507) | (0.0458) | (0.0653) |
| **Religion (base cat: Hindu)** | | | | |
| Muslims | -0.0610** | 0.0136 | -0.118** | -0.0588 |
| | (0.0256) | (0.0478) | (0.0464) | (0.0418) |
| Christians | 0.0287 | 0.0395 | -0.0429 | 0.00522 |
| | (0.0300) | (0.0489) | (0.0392) | (0.0828) |
| Others | 0.0640 | 0.122 | -0.0204 | 0.110 |
| | (0.0465) | (0.0836) | (0.0679) | (0.0690) |
| Other controls | | | | |
| Sector (rural/urban) | Y | Y | Y | Y |
| Origin location (state) | Y | N | N | Y |
| Destination location (state) | Y | Y | Y | Y |
| Industry category (1 digit) | Y | Y | Y | Y |
| Occupation category (1 digit) | Y | Y | Y | Y |
| Reason for migration | Y | Y | Y | Y |
| Migration stream | Y | Y | Y | Y |
| Migration distance | Y | N | N | N |
| Years since migration | Y | Y | Y | Y |
| International immigration rate (state level) | Y | Y | Y | Y |
| ln(working age population) | Y | Y | Y | Y |
| ln(labour force) | Y | Y | Y | Y |
| Employment status (before migration) | Y | Y | Y | Y |
| **Sample selection corrections** | | | | |
| Inverse mills ratio- employment | Y | Y | Y | Y |
| Inverse mills ratio- occupation | Y | Y | Y | Y |
| Inverse mills ratio- migration | Y | Y | Y | Y |
| | | | | |
| Clustered standard errors (at first stage unit) | Y | Y | Y | Y |
| Observations | 13,745 | 4,518 | 4,953 | 4,274 |
| R-squared | 0.512 | 0.540 | 0.541 | 0.464 |

Source: Authors' calculation based on NSSO employment, unemployment and migration survey, 2007-08.
Note:
(i) *** signals significant at 1% level, ** signals significant at 5% level and * signals significant at 10% level.
(ii) robust standard errors are given in parenthesis.

Table A9a: Returns to education and EOM for wage/salary employed migrants in working-age group: by years since migration

| | Dependent variable: ln(wage) | | | | |
|---|---|---|---|---|---|
| | (1) | (2) | (3) | (4) | (5) |
| Explanatory variables | Overall | 0-2 | 3-5 | 6-10 | 11 and above |
| Attained education | 0.0298*** | 0.0303*** | 0.0276*** | 0.0240*** | 0.0315*** |
| | (0.0018) | (0.0031) | (0.00375) | (0.0037) | (0.0034) |
| Age | 0.0108** | 0.00714 | 0.0322** | 0.0225* | 0.0305* |
| | (0.0052) | (0.00835) | (0.0125) | (0.0133) | (0.0156) |
| Age squared | 3.16e-06 | 4.46e-05 | -0.0003** | -0.000215 | -0.0002 |
| | (6.50e-05) | (0.0001) | (0.0002) | (0.0002) | (0.0002) |
| **Gender (base cat: male)** | | | | | |
| Female | -0.206*** | -0.0674 | -0.202* | -0.339** | -0.332** |
| | (0.0507) | (0.0668) | (0.109) | (0.162) | (0.154) |
| **Marital status (base cat: unmarried male)** | | | | | |
| Male*married | 0.0948*** | 0.0587 | 0.0223 | 0.0796 | 0.184** |
| | (0.0248) | (0.0359) | (0.0494) | (0.0577) | (0.0852) |
| Male*Others | 0.0448 | -0.0240 | 0.0102 | -0.0738 | 0.154 |
| | (0.0544) | (0.106) | (0.0884) | (0.118) | (0.118) |
| **Gender*marital status** | | | | | |
| Female*married | -0.203*** | -0.253** | -0.118 | -0.105 | -0.585*** |
| | (0.0710) | (0.105) | (0.138) | (0.212) | (0.196) |
| Female*others | -0.160* | -0.200 | -0.201 | 0.109 | -0.347* |
| | (0.0868) | (0.147) | (0.149) | (0.229) | (0.201) |
| **Social group (base cat: Scheduled tribe)** | | | | | |
| Scheduled caste | -0.0343 | -0.0352 | -0.0196 | -0.107** | 0.00650 |
| | (0.0297) | (0.0459) | (0.0666) | (0.0518) | (0.0625) |
| Other Backward Class (OBC) | -0.0165 | 0.00469 | -0.0642 | -0.0588 | 0.0207 |
| | (0.0287) | (0.0455) | (0.0666) | (0.0511) | (0.0616) |
| Others | 0.0264 | 0.0636 | 0.00704 | -0.0354 | 0.0264 |
| | (0.0297) | (0.0465) | (0.0668) | (0.0536) | (0.0648) |
| **Religion (base cat: Hindu)** | | | | | |
| Muslims | -0.0512** | -0.0185 | -0.0437 | -0.0774 | -0.0665 |
| | (0.0256) | (0.0411) | (0.0493) | (0.0622) | (0.0453) |
| Christians | 0.0171 | -0.103** | 0.0517 | 0.0288 | 0.126** |
| | (0.0311) | (0.0489) | (0.0642) | (0.0740) | (0.0556) |
| Others | 0.0541 | 0.161*** | 0.0650 | -0.0765 | -0.0137 |
| | (0.0478) | (0.0602) | (0.124) | (0.133) | (0.0732) |
| **Other controls** | | | | | |
| Sector (rural/urban) | Y | Y | Y | Y | Y |
| Origin location (state) | Y | Y | Y | Y | Y |
| Destination location (state) | Y | Y | Y | Y | Y |
| Industry category (1 digit) | Y | Y | Y | Y | Y |
| Occupation category (1 digit) | Y | Y | Y | Y | Y |
| Reason for migration | Y | Y | Y | Y | Y |
| Migration stream | Y | Y | Y | Y | Y |
| Migration distance | Y | Y | Y | Y | Y |
| Years since migration | Y | N | N | N | N |
| International immigration rate (state level) | Y | Y | Y | Y | Y |
| ln(working age population) | Y | Y | Y | Y | Y |
| ln(labour force) | Y | Y | Y | Y | Y |
| Employment status (before migration) | Y | Y | Y | Y | Y |
| **Sample selection corrections** | | | | | |
| Inverse mills ratio- employment | Y | Y | Y | Y | Y |
| Inverse mills ratio- occupation | Y | Y | Y | Y | Y |
| Inverse mills ratio- migration | Y | Y | Y | Y | Y |
| | | | | | |
| Clustered standard errors (at first stage unit) | Y | Y | Y | Y | Y |
| Observations | 13,887 | 4,714 | 2,875 | 2,553 | 3,745 |
| R-squared | 0.501 | 0.506 | 0.532 | 0.548 | 0.506 |

Source: Authors' calculation based on NSSO employment, unemployment and migration survey, 2007-08.
Note:
(i) *** signals significant at 1% level, ** signals significant at 5% level and * signals significant at 10% level.
(ii) robust standard errors are given in parenthesis.

Table A9b: Returns to education and EOM for wage/salary employed migrants in working-age group: by years since migration

| | Dependent variable: ln(wage) | | | | |
|---|---|---|---|---|---|
| | (1) | (2) | (3) | (4) | (5) |
| | | | | | 11 and |
| Explanatory variables | Overall | 0-2 | 3-5 | 6-10 | above |
| Required education | 0.075*** | 0.062*** | 0.064*** | 0.073*** | 0.094*** |
| | (0.0058) | (0.0096) | (0.0116) | (0.0119) | (0.0113) |
| Surplus education | 0.032*** | 0.034*** | 0.019** | 0.03*** | 0.042*** |
| | (0.003) | (0.0053) | (0.0079) | (0.0071) | (0.0067) |
| Deficit education | -0.02*** | -0.02*** | -0.03*** | -0.02** | -0.02*** |
| | (0.0036) | (0.006) | (0.009) | (0.007) | (0.005) |
| Age | 0.0106** | 0.00831 | 0.029** | 0.0221* | 0.0327** |
| | (0.0052) | (0.008) | (0.0124) | (0.0133) | (0.0150) |
| Age squared | 5.82e-07 | 3.08e-05 | -0.00028* | -0.000213 | -0.000240 |
| | (6.46e-05) | (0.0001) | (0.0002) | (0.00017) | (0.0002) |
| **Gender (base cat: male)** | | | | | |
| Female | -0.192*** | -0.0576 | -0.194* | -0.343** | -0.337** |
| | (0.0504) | (0.0668) | (0.110) | (0.157) | (0.170) |
| **Marital status (base cat: unmarried male)** | | | | | |
| Male*married | 0.0968*** | 0.0614* | 0.0349 | 0.0703 | 0.163* |
| | (0.0246) | (0.0356) | (0.0486) | (0.0571) | (0.0851) |
| Male*Others | 0.0433 | -0.0235 | 0.00498 | -0.0931 | 0.142 |
| | (0.0537) | (0.112) | (0.0849) | (0.119) | (0.116) |
| **Gender*marital status** | | | | | |
| Female*married | -0.167** | -0.228** | -0.0891 | -0.0384 | -0.467** |
| | (0.0703) | (0.104) | (0.139) | (0.209) | (0.208) |
| Female*others | -0.125 | -0.183 | -0.145 | 0.189 | -0.263 |
| | (0.0859) | (0.151) | (0.147) | (0.226) | (0.211) |
| **Social group (base cat: Scheduled tribe)** | | | | | |
| Scheduled caste | -0.0272 | -0.0332 | -0.0121 | -0.0974* | 0.0223 |
| | (0.0295) | (0.0457) | (0.0663) | (0.0517) | (0.0625) |
| Other Backward Class (OBC) | -0.0128 | 0.0131 | -0.0657 | -0.0495 | 0.0305 |
| | (0.0289) | (0.0454) | (0.0680) | (0.0500) | (0.0617) |
| Others | 0.0300 | 0.0654 | 0.00343 | -0.0225 | 0.0460 |
| | (0.0298) | (0.0463) | (0.0689) | (0.0526) | (0.0644) |
| **Religion (base cat: Hindu)** | | | | | |
| Muslims | -0.0610** | -0.0184 | -0.0558 | -0.0890 | -0.0854* |
| | (0.0256) | (0.0411) | (0.0500) | (0.0629) | (0.0444) |
| Christians | 0.0287 | -0.0860* | 0.0654 | 0.0407 | 0.124** |
| | (0.0300) | (0.0455) | (0.0640) | (0.0709) | (0.0543) |
| Others | 0.0640 | 0.151*** | 0.0523 | -0.0664 | 0.0472 |
| | (0.0465) | (0.0571) | (0.123) | (0.135) | (0.0606) |
| **Other controls** | | | | | |
| Sector (rural/urban) | Y | Y | Y | Y | Y |
| Origin location (state) | Y | Y | Y | Y | Y |
| Destination location (state) | Y | Y | Y | Y | Y |
| Industry category (1 digit) | Y | Y | Y | Y | Y |
| Occupation category (1 digit) | Y | Y | Y | Y | Y |
| Reason for migration | Y | Y | Y | Y | Y |
| Migration stream | Y | Y | Y | Y | Y |
| Migration distance | Y | Y | Y | Y | Y |
| Years since migration | Y | N | N | N | N |
| International immigration rate (state level) | Y | Y | Y | Y | Y |
| ln(working age population) | Y | Y | Y | Y | Y |
| ln(labour force) | Y | Y | Y | Y | Y |
| Employment status (before migration) | Y | Y | Y | Y | Y |
| **Sample selection corrections** | | | | | |
| Inverse mills ratio- employment | Y | Y | Y | Y | Y |
| Inverse mills ratio- occupation | Y | Y | Y | Y | Y |
| Inverse mills ratio- migration | Y | Y | Y | Y | Y |
| | | | | | |
| Clustered standard errors (at first stage unit) | Y | Y | Y | Y | Y |
| Observations | 13,745 | 4,670 | 2,851 | 2,528 | 3,696 |
| R-squared | 0.512 | 0.511 | 0.536 | 0.553 | 0.531 |

Source: Authors' calculation based on NSSO employment, unemployment and migration survey, 2007-08.
Note:
(i) *** signals significant at 1% level, ** signals significant at 5% level and * signals significant at 10% level.
(ii) robust standard errors are given in parenthesis.

Table A10: Returns to education for wage/salary employed in working-age group: Work-related migrants and non-migrants

| | Dependent variable: ln(wage) | | | |
|---|---|---|---|---|
| | (1) | (2) | (3) | (4) |
| | Non migrants | | Intra-district migrants | |
| Explanatory variables | OLS | Lewbel IV | OLS | Lewbel IV |
| Attained education | 0.0232*** | 0.0219*** | 0.0291*** | 0.0396*** |
| | (0.000922) | (0.00405) | (0.00325) | (0.00777) |
| Age | 0.00308 | 0.00148 | 0.00638 | 0.0114 |
| | (0.00279) | (0.00286) | (0.00926) | (0.00935) |
| Age squared | 8.27e-05** | 0.000102*** | 5.44e-05 | 1.78e-05 |
| | (3.58e-05) | (3.83e-05) | (0.000116) | (0.000116) |
| **Gender (base cat: male)** | | | | |
| Female | -0.242*** | -0.260*** | -0.214** | -0.273** |
| | (0.0243) | (0.0262) | (0.107) | (0.106) |
| **Marital status (base cat: unmarried male)** | | | | |
| Male*married | 0.0843*** | 0.0895*** | 0.218*** | 0.257*** |
| | (0.0142) | (0.0137) | (0.0580) | (0.0552) |
| Male*Others | -0.0235 | -0.0278 | 0.206** | 0.236*** |
| | (0.0248) | (0.0257) | (0.0817) | (0.0829) |
| **Gender*marital status** | | | | |
| Female*married | 0.0734*** | 0.0793*** | -0.313** | 0.0213 |
| | (0.0249) | (0.0257) | (0.144) | (0.0935) |
| Female*others | 0.0310 | 0.0480 | -0.145 | 0.136 |
| | (0.0302) | (0.0299) | (0.143) | (0.104) |
| **Social group (base cat: Scheduled tribe)** | | | | |
| Scheduled caste | 0.0165 | 0.0190* | -0.0102 | -0.0173 |
| | (0.0150) | (0.0112) | (0.0442) | (0.0435) |
| Other Backward Class (OBC) | 0.0252* | 0.0271** | 0.00927 | 0.0220 |
| | (0.0151) | (0.0114) | (0.0463) | (0.0444) |
| Others | 0.0985*** | 0.0936*** | 0.0456 | 0.0577 |
| | (0.0174) | (0.0145) | (0.0501) | (0.0478) |
| **Religion (base cat: Hindu)** | | | | |
| Muslims | -0.0272** | -0.0254** | 0.0323 | -0.00923 |
| | (0.0126) | (0.0129) | (0.0484) | (0.0417) |
| Christians | 0.0639** | 0.0491** | 0.0282 | 0.0584 |
| | (0.0271) | (0.0221) | (0.0517) | (0.0518) |
| Others | 0.0507* | 0.0494** | 0.0860 | 0.0858 |
| | (0.0270) | (0.0219) | (0.0892) | (0.0878) |
| **Other controls** | | | | |
| Sector (rural/urban) | Y | Y | Y | Y |
| Origin location (state) | N | N | Y | N |
| Destination location (state) | Y | Y | Y | Y |
| Industry (1 digit) | Y | Y | Y | Y |
| Occupation (1 digit) | Y | Y | Y | Y |
| Reason for migration | N | N | Y | N |
| Migration stream | N | N | Y | N |
| Migration distance | N | N | Y | N |
| Years since migration | N | N | Y | N |
| International immigration rate (district level) | Y | Y | Y | Y |
| ln(working age population) | Y | Y | Y | Y |
| ln(labour force) | Y | Y | Y | Y |
| Employment status (before migration) | N | N | Y | Y |
| **Sample selection corrections** | | | | |
| Inverse mills ratio- employment | Y | Y | Y | Y |
| Inverse mills ratio- occupation | Y | Y | Y | Y |
| Inverse mills ratio- migration | N | N | Y | Y |
| | | | | |
| Clustered standard errors (at first stage unit) | Y | Y | Y | Y |
| Observations | 56,440 | 54,320 | 4,592 | 4,516 |
| R-squared | 0.357 | 0.351 | 0.528 | 0.501 |

Source: Authors' calculation based on NSSO employment, unemployment and migration survey, 2007-08.
Note: (i) *** signals significant at 1% level ,** signals significant at 5% level and * signals significant at 10% level. (ii) Standard errors are given in parenthesis.

## Table A11: Sensitivity check of incidence of EOM

### Threshold of 1 standard deviation from mean

| Match status | Non-migrants | Work related migrants | Total |
|---|---|---|---|
| Undereducated | 11.71 | 15.04 | 12.12 |
| Adequately educated | 70.5 | 70.32 | 70.48 |
| Overeducated | 17.79 | 14.64 | 17.4 |
| Total | 100 | 100 | 100 |

### Threshold of 0.9 standard deviation from mean

| Match status | Non-migrants | Work related migrants | Total |
|---|---|---|---|
| Undereducated | 12.66 | 16.86 | 13.18 |
| Adequately educated | 59.68 | 63.35 | 60.13 |
| Overeducated | 27.67 | 19.79 | 26.69 |
| Total | 100 | 100 | 100 |

### Threshold of 1.1 standard deviation from mean

| Match status | Non-migrants | Work related migrants | Total |
|---|---|---|---|
| Undereducated | 6.75 | 10.64 | 7.23 |
| Adequately educated | 78.39 | 78.39 | 78.39 |
| Overeducated | 14.86 | 10.98 | 14.38 |
| Total | 100 | 100 | 100 |

### Threshold of 1 standard deviation from median

| Match status | Non-migrants | Work related migrants | Total |
|---|---|---|---|
| Undereducated | 12.81 | 18.56 | 13.53 |
| Adequately educated | 55.98 | 67.01 | 57.35 |
| Overeducated | 31.21 | 14.42 | 29.12 |
| Total | 100 | 100 | 100 |

Source: Authors' calculation based on NSSO employment, unemployment and migration survey, 2007-08.
*Note*: Sampling weights have been used.

| Table A12: Diagnostic tests for Lewbel IV models | | |
|---|---|---|
| Test | Non-migrants | Intra-district migrants |
| Breusch-Pagan / Cook-Weisberg test for heteroskedasticity | 2221.62 (0.00) | 135.45 (0.00) |
| Weak Instrument test (Sanderson-Windmeijer (SW) test) | 2221.1(0.00) | 983.86 (0.00) |
| Under identification test (Kleibergen- Papp rk LM statistics) | 3156.08(0.00) | 929.14 (0.00) |
| Weak identification test (Cragg-Donald Wald F statistics) | 80.218 (0.00) | 17.989(0.00) |
| Hansen J Statistics | 1501.06 (0.00) | 320.473 (0.00) |

Source: Authors' calculation.
Note: p-values are in parenthesis.

Table A13a: Returns to education and EOM for wage/salary employed migrants in working-age group: by reason for migration

| | (1) | (2) | (3) | (4) | (5) | (6) | (7) |
|---|---|---|---|---|---|---|---|
| | | | | Dependent variable: ln(wage) | | | |
| | | Work | | Forced | | | |
| Explanatory variables | All | related | Education | migration | Marriage | Tied Movers | Others |
| Attained education | 0.030*** | 0.0298*** | 0.0644 | 0.0336*** | 0.0133*** | 0.0192*** | 0.037*** |
| | (0.0013) | (0.00182) | (0.0544) | (0.0100) | (0.00238) | (0.00430) | (0.0077) |
| Age | 0.0084** | 0.0108** | 0.0466 | 0.0108 | 0.0466*** | 0.0493*** | 0.0524** |
| | (0.0042) | (0.00519) | (0.0904) | (0.0335) | (0.00941) | (0.0147) | (0.0251) |
| Age squared | 2.54e-05 | 3.16e-06 | -0.000306 | 1.53e-05 | -0.000523*** | -0.000539*** | -0.000530* |
| | (5.25e-05) | (6.50e-05) | (0.00110) | (0.000416) | (0.000117) | (0.000194) | (0.000315) |
| **Gender (base cat: male)** | | | | | | | |
| Female | -0.149*** | -0.206*** | -0.764 | -0.0255 | -0.364** | -0.239*** | -0.135 |
| | (0.0366) | (0.0507) | (0.556) | (0.287) | (0.153) | (0.0808) | (0.197) |
| **Marital status (base cat: unmarried male)** | | | | | | | |
| Male*married | 0.116*** | 0.0948*** | 0.553 | 0.199 | -0.194* | 0.143*** | 0.0550 |
| | (0.0180) | (0.0248) | (0.475) | (0.150) | (0.111) | (0.0535) | (0.101) |
| Male*Others | -0.0358 | 0.0448 | 0.977 | 0.248 | -0.681*** | -0.134 | -0.0246 |
| | (0.0506) | (0.0544) | (1.370) | (0.259) | (0.173) | (0.116) | (0.168) |
| **Gender*marital status** | | | | | | | |
| Female*married | -0.387*** | -0.203*** | -0.461 | -0.516 | -0.394** | -0.342** | -0.771** |
| | (0.0580) | (0.0710) | (1.156) | (0.438) | (0.194) | (0.158) | (0.349) |
| Female*others | -0.206*** | -0.160* | | -0.697 | 0.261 | 0.192 | -0.772** |
| | (0.0729) | (0.0868) | | (0.466) | (0.223) | (0.189) | (0.386) |
| **Social group (base cat: Scheduled tribe)** | | | | | | | |
| Scheduled caste | -0.0459** | -0.0343 | 0.899* | -0.130 | -0.0136 | -0.0579 | -0.169* |
|  | (0.0190) | (0.0297) | (0.494) | (0.136) | (0.0254) | (0.0625) | (0.0920) |
| Other Backward Class (OBC) | -0.0591*** | -0.0165 | 0.549 | -0.0156 | -0.0424 | -0.0501 | -0.169* |
|  | (0.0187) | (0.0287) | (0.447) | (0.118) | (0.0260) | (0.0627) | (0.0994) |
| Others | -0.000308 | 0.0264 | 0.320 | 0.0550 | -0.00515 | 0.135** | -0.0982 |
|  | (0.0211) | (0.0297) | (0.446) | (0.124) | (0.0371) | (0.0687) | (0.105) |
| **Religion (base cat: Hindu)** |  |  |  |  |  |  |  |
| Muslims | -0.0303 | -0.0512** | 0.617 | 0.216* | -0.0269 | -0.144** | 0.0461 |
|  | (0.0203) | (0.0256) | (0.520) | (0.120) | (0.0381) | (0.0576) | (0.115) |
| Christians | 0.0238 | 0.0171 | -0.485 | 0.0158 | 0.0389 | -0.0124 | -0.00597 |
|  | (0.0245) | (0.0311) | (0.653) | (0.138) | (0.0541) | (0.0662) | (0.0879) |
| Others | -0.0307 | 0.0541 | 0.946 | 0.240* | -0.112*** | -0.145** | -0.313* |
|  | (0.0311) | (0.0478) | (0.605) | (0.134) | (0.0425) | (0.0710) | (0.174) |
| **Other controls** |  |  |  |  |  |  |  |
| Sector (rural/urban) | Y | Y | Y | Y | Y | Y | Y |
| Origin location (state) | Y | Y | Y | Y | Y | Y | Y |
| Destination location (state) | Y | Y | Y | Y | Y | Y | Y |
| Industry category (1 digit) | Y | Y | Y | Y | Y | Y | Y |
| Occupation category (1 digit) | Y | Y | Y | Y | Y | Y | Y |
| Reason for migration | Y | N | N | N | N | N | N |
| Migration stream | Y | Y | Y | Y | Y | Y | Y |
| Migration distance | Y | Y | Y | Y | Y | Y | Y |
| Years since migration | Y | Y | Y | Y | Y | Y | Y |
| International immigration rate (state level) | Y | Y | Y | Y | Y | Y | Y |
| ln(working age population) | Y | Y | Y | Y | Y | Y | Y |
| ln(labour force) | Y | Y | Y | Y | Y | Y | Y |
| employment status (before migration) | Y | Y | Y | Y | Y | Y | Y |
| **Sample selection corrections** |  |  |  |  |  |  |  |
| Inverse mills ratio- employment | Y | Y | Y | Y | Y | Y | Y |
| Inverse mills ratio- occupation | Y | Y | Y | Y | Y | Y | Y |
| Inverse mills ratio- migration | Y | Y | Y | Y | Y | Y | Y |
| Clustered standard errors (at first stage unit) | Y | Y | Y | Y | Y | Y | Y |
| Observations | 27,896 | 13,887 | 147 | 652 | 9,482 | 2,880 | 804 |
| R-squared | 0.565 | 0.501 | 0.765 | 0.591 | 0.347 | 0.443 | 0.515 |

Source: Authors' calculation based on NSSO employment, unemployment and migration survey, 2007-08.
Note:
(i) *** signals significant at 1% level, ** signals significant at 5% level and * signals significant at 10% level.
(ii) robust standard errors are given in parenthesis.

Table A13b: Returns to education and EOM for wage/salary employed migrants in working-age group: by reason for migration

| | Dependent variable: ln(wage) | | | | | | |
|---|---|---|---|---|---|---|---|
| | (1) | (2) | (3) | (4) | (5) | (6) | (7) |
| | | Work related | | Forced migration | | Tied Movers | |
| Explanatory variables | All | | Education | | Marriage | | Others |
| Required education | 0.104*** | 0.0751*** | 0.275** | 0.103** | 0.175*** | 0.0492** | 0.0976*** |
| | (0.00535) | (0.00584) | (0.131) | (0.0486) | (0.0161) | (0.0197) | (0.0322) |
| Surplus education | 0.0263*** | 0.0324*** | 0.0711 | 0.0198 | -0.00231 | 0.0163* | 0.0215* |
| | (0.00267) | (0.00334) | (0.0732) | (0.0191) | (0.00520) | (0.00876) | (0.0125) |
| Deficit education | -0.0267*** | -0.0241*** | -0.0161 | -0.0407** | -0.0248*** | -0.0182** | -0.0554*** |
| | (0.00276) | (0.00357) | (0.142) | (0.0178) | (0.00556) | (0.00772) | (0.0191) |
| Age | 0.00906** | 0.0106** | 0.0409 | 0.0117 | 0.0419*** | 0.0474*** | 0.0547** |
| | (0.00416) | (0.00514) | (0.0842) | (0.0329) | (0.00931) | (0.0151) | (0.0241) |
| Age squared | 1.08e-05 | 5.82e-07 | -0.000232 | -1.63e-05 | -0.000469*** | -0.000510** | -0.000575* |
| | (5.23e-05) | (6.46e-05) | (0.00102) | (0.000408) | (0.000115) | (0.000198) | (0.000304) |
| **Gender (base cat: male)** | | | | | | | |
| Female | -0.139*** | -0.192*** | -0.513 | 0.0583 | -0.0997 | -0.241*** | -0.117 |
| | (0.0365) | (0.0504) | (0.552) | (0.279) | (0.187) | (0.0818) | (0.188) |
| **Marital status (base cat: unmarried male)** | | | | | | | |
| Male*married | 0.126*** | 0.0968*** | 0.353 | 0.252 | -0.0422 | 0.156*** | 0.0510 |
| | (0.0181) | (0.0246) | (0.497) | (0.153) | (0.155) | (0.0543) | (0.0988) |
| Male*Others | -0.0311 | 0.0433 | 0.390 | 0.268 | -0.486** | -0.122 | -0.0872 |
| | (0.0500) | (0.0537) | (1.369) | (0.247) | (0.202) | (0.118) | (0.143) |
| **Gender*marital status** | | | | | | | |
| Female*married | -0.318*** | -0.167** | -0.482 | -0.552 | -0.479** | -0.328** | -0.560 |
| | (0.0577) | (0.0703) | (1.272) | (0.447) | (0.216) | (0.159) | (0.346) |
| Female*others | -0.131* | -0.125 | | -0.700 | 0.114 | 0.193 | -0.573 |
| | (0.0723) | (0.0859) | | (0.473) | (0.242) | (0.191) | (0.369) |

**Social group (base cat: Scheduled tribe)**

| | (1) | (2) | (3) | (4) | (5) | (6) | (7) |
|---|---|---|---|---|---|---|---|
| Scheduled caste | -0.0311 | -0.0272 | 0.774 | -0.113 | 0.00533 | -0.0432 | -0.123 |
| | (0.0191) | (0.0295) | (0.519) | (0.142) | (0.0257) | (0.0630) | (0.0911) |
| Other Backward Class (OBC) | -0.0424** | -0.0128 | 0.443 | 0.000472 | -0.0168 | -0.0244 | -0.135 |
| | (0.0188) | (0.0289) | (0.488) | (0.118) | (0.0263) | (0.0624) | (0.102) |
| Others | 0.0177 | 0.0300 | 0.0997 | 0.0832 | 0.0199 | 0.157** | -0.0222 |
| | (0.0212) | (0.0298) | (0.460) | (0.124) | (0.0373) | (0.0681) | (0.103) |

**Religion (base cat: Hindu)**

| | (1) | (2) | (3) | (4) | (5) | (6) | (7) |
|---|---|---|---|---|---|---|---|
| Muslims | -0.0372* | -0.0610** | 0.869 | 0.228* | -0.0384 | -0.142** | 0.0401 |
| | (0.0202) | (0.0256) | (0.575) | (0.126) | (0.0378) | (0.0582) | (0.112) |
| Christians | 0.0338 | 0.0287 | -0.405 | 0.0368 | 0.0337 | 0.00466 | -0.0208 |
| | (0.0240) | (0.0300) | (0.731) | (0.147) | (0.0530) | (0.0692) | (0.0940) |
| Others | -0.0216 | 0.0640 | 0.777 | 0.284** | -0.0940** | -0.154** | -0.313* |
| | (0.0307) | (0.0465) | (0.612) | (0.141) | (0.0440) | (0.0708) | (0.176) |

**Other controls**

| | (1) | (2) | (3) | (4) | (5) | (6) | (7) |
|---|---|---|---|---|---|---|---|
| Sector (rural/urban) | Y | Y | Y | Y | Y | Y | Y |
| Origin location (state) | Y | Y | Y | Y | Y | Y | Y |
| Destination location (state) | Y | Y | Y | Y | Y | Y | Y |
| Industry category (1 digit) | Y | Y | Y | Y | Y | Y | Y |
| Occupation category (1 digit) | Y | Y | Y | Y | Y | Y | Y |
| Reason for migration | Y | N | N | N | N | N | N |
| Migration stream | Y | Y | Y | Y | Y | Y | Y |
| Migration distance | Y | Y | Y | Y | Y | Y | Y |
| Years since migration | Y | N | N | N | N | N | N |
| International immigration rate (district level) | Y | Y | Y | Y | Y | Y | Y |
| ln(working age population) | Y | Y | Y | Y | Y | Y | Y |
| ln(labour force) | Y | Y | Y | Y | Y | Y | Y |
| employment status (before migration) | Y | Y | Y | Y | Y | Y | Y |

**Sample selection corrections**

| | | | | | | | |
|---|---|---|---|---|---|---|---|
| Inverse mills ratio- employment | Y | Y | Y | Y | Y | Y | Y |
| Inverse mills ratio- occupation | Y | Y | Y | Y | Y | Y | Y |
| Inverse mills ratio- migration | Y | Y | Y | Y | Y | Y | Y |
| | | | | | | | |
| Clustered standard errors (at first stage unit) | Y | Y | Y | Y | Y | Y | Y |
| Observations | 27,225 | 13,745 | 147 | 638 | 9,059 | 2,823 | 773 |
| R-squared | 0.578 | 0.512 | 0.782 | 0.597 | 0.376 | 0.447 | 0.559 |

Source: Authors' calculation based on NSSO employment, unemployment and migration survey, 2007-08.
Note:
(i) *** signals significant at 1% level, ** signals significant at 5% level and * signals significant at 10% level.
(ii) robust standard errors are given in parenthesis.

Table A14a: Returns to education and EOM for wage/salary employed married individuals in working-age group

| Explanatory variables | Dependent variable: ln(wage) | |
|---|---|---|
| | (1) Work related migrant | (2) Non-migrant |
| Attained education | 0.0225*** | 0.0162*** |
| | (0.00248) | (0.00109) |
| Spouse's education (in years) | 0.0113*** | 0.00838*** |
| | (0.00192) | (0.00111) |
| **Spouse's work status (base cat: not employed)** | | |
| Employed | -0.113*** | -0.0610*** |
| | (0.0182) | (0.0104) |
| Age | 0.0140** | 0.00677** |
| | (0.00715) | (0.00343) |
| Age squared | -3.86e-05 | 4.58e-05 |
| | (8.75e-05) | (4.42e-05) |
| **Gender (base cat: male)** | | |
| Female | -0.638*** | -0.850*** |
| | (0.102) | (0.0518) |
| **Social group (base cat: Scheduled tribe)** | | |
| Scheduled caste | -0.0166 | 0.00591 |
| | (0.0344) | (0.0160) |
| Other Backward Class (OBC) | -0.0436 | -0.00940 |
| | (0.0327) | (0.0165) |
| Others | -0.00196 | 0.0393** |
| | (0.0342) | (0.0199) |
| **Religion (base cat: Hindu)** | | |
| Muslims | -0.0831** | 0.0367** |
| | (0.0344) | (0.0170) |
| Christians | 0.0569 | 0.0565* |
| | (0.0373) | (0.0326) |
| Others | -0.0180 | 0.0535* |
| | (0.0539) | (0.0301) |
| **Other controls** | | |
| Sector (rural/urban) | Y | Y |
| Origin location (state) | Y | N |
| Destination location (state) | Y | Y |
| Industry category (1 digit) | Y | Y |
| Occupation category (1 digit) | Y | Y |
| Reason for migration | Y | N |
| Migration stream | Y | N |
| Migration distance | Y | N |
| Years since migration | Y | N |
| International immigration rate (state level) | Y | Y |
| ln(working age population) | Y | Y |
| ln(labour force) | Y | Y |
| employment status (before migration) | Y | N |
| **Sample selection corrections** | | |
| Inverse mills ratio- employment | Y | Y |
| Inverse mills ratio- occupation | Y | Y |
| Inverse mills ratio- migration | Y | N |
| | | |
| Clustered standard errors (at first stage unit) | Y | Y |
| Observations | 8,551 | 36,736 |
| R-squared | 0.521 | 0.410 |

Source: Authors' calculation based on NSSO employment, unemployment and migration survey, 2007-08.
Note:
(i) *** signals significant at 1% level, ** signals significant at 5% level and * signals significant at 10% level.
(ii) robust standard errors are given in parenthesis.

Table A14b: Returns to education and EOM for wage/salary employed married individuals in working-age group

| Explanatory variables | Dependent variable: ln(wage) | |
| | (1) Work related migrant | (2) Non-migrant |
| --- | --- | --- |
| Required education | 0.0684*** | 0.0769*** |
| | (0.00682) | (0.00555) |
| Surplus education | 0.0236*** | 0.0103*** |
| | (0.00419) | (0.00231) |
| Deficit education | -0.0198*** | -0.0195*** |
| | (0.00524) | (0.00225) |
| Spouse's education (in years) | 0.0102*** | 0.00819*** |
| | (0.00189) | (0.00113) |
| **Spouse's work status (base cat: not employed)** | | |
| Employed | -0.105*** | -0.0616*** |
| | (0.0179) | (0.0103) |
| Age | 0.0130* | 0.00935*** |
| | (0.00701) | (0.00343) |
| Age squared | -3.72e-05 | 1.18e-05 |
| | (8.59e-05) | (4.41e-05) |
| **Gender (base cat: male)** | | |
| Female | -0.562*** | -0.801*** |
| | (0.0996) | (0.0521) |
| **Social group (base cat: Scheduled tribe)** | | |
| Scheduled caste | -0.00615 | 0.0138 |
| | (0.0343) | (0.0161) |
| Other Backward Class (OBC) | -0.0382 | -0.00500 |
| | (0.0328) | (0.0168) |
| Others | 0.00537 | 0.0421** |
| | (0.0342) | (0.0202) |
| **Religion (base cat: Hindu)** | | |
| Muslims | -0.101*** | 0.0307* |
| | (0.0343) | (0.0172) |
| Christians | 0.0511 | 0.0594* |
| | (0.0372) | (0.0339) |
| Others | 0.00664 | 0.0436 |
| | (0.0512) | (0.0304) |
| **Other controls** | | |
| Sector (rural/urban) | Y | Y |
| Origin location (state) | Y | N |
| Destination location (state) | Y | Y |
| Industry category (1 digit) | Y | Y |
| Occupation category (1 digit) | Y | Y |
| Reason for migration | Y | N |
| Migration stream | Y | N |
| Migration distance | Y | N |
| Years since migration | Y | N |
| International immigration rate (state level) | Y | Y |
| ln(working age population) | Y | Y |
| ln(labour force) | Y | Y |
| employment status (before migration) | Y | N |
| **Sample selection corrections** | | |
| Inverse mills ratio- employment | Y | Y |
| Inverse mills ratio- occupation | Y | Y |
| Inverse mills ratio- migration | Y | N |
| | | |
| Clustered standard errors (at first stage unit) | Y | Y |
| Observations | 8,440 | 34,968 |
| R-squared | 0.534 | 0.429 |

Source: Authors' calculation based on NSSO employment, unemployment and migration survey, 2007-08.
Note:
 (i) *** signals significant at 1% level, ** signals significant at 5% level and * signals significant at 10% level.
 (ii) robust standard errors are given in parenthesis.

Table A15: Returns to education and EOM for wage/salary employed migrants in working-age group: by gender

| Explanatory variables | Dependent variable: ln(wage) | | | |
| | (1) Overall | (2) Male | (3) Female | Chow test |
|---|---|---|---|---|
| Attained education | 0.0298*** | 0.0292*** | 0.0371*** | 1.8 |
| | (0.00182) | (0.00191) | (0.00584) | |
| Observations | 13,887 | 12,301 | 1,586 | |
| R-squared | 0.501 | 0.477 | 0.635 | |
| Required education | 0.0751*** | 0.0706*** | 0.106*** | 2.97* |
| | (0.00584) | (0.00617) | (0.0202) | |
| Surplus education | 0.0324*** | 0.0324*** | 0.0376*** | 0.19 |
| | (0.00334) | (0.00348) | (0.0118) | |
| Deficit education | -0.0241*** | -0.0236*** | -0.0317*** | 0.51 |
| | (0.00357) | (0.00378) | (0.0112) | |
| Observations | 13,745 | 12,173 | 1,572 | |
| R-squared | 0.512 | 0.489 | 0.639 | |

Source: Authors' calculation based on NSSO employment, unemployment and migration survey, 2007-08.
Note:
(i) *** signals significant at 1% level, ** signals significant at 5% level and * signals significant at 10% level.
(ii) robust standard errors are given in parenthesis.
(iii) Chow test indicates whether the difference in the coefficients is significant or not.